\documentclass[%
 reprint,
superscriptaddress,
preprintnumbers,
 amsmath,amssymb,
 aps,nofootinbib
]{revtex4-1}
\usepackage{dsfont}
\usepackage{graphicx}
\usepackage{dcolumn}
\usepackage{bm}
\usepackage{color}
\usepackage{epsfig}

\usepackage[colorlinks]{hyperref}
\definecolor{linkcolor}{rgb}{0.0,0.5,0.4}
\hypersetup{
     colorlinks   = true,
     citecolor    = linkcolor,
     linkcolor    = linkcolor,
     urlcolor    = linkcolor
}

\usepackage{natbib}
\usepackage{float}
\usepackage{footmisc}
\usepackage{slashed}
\definecolor{lightred}{rgb}{1,0.5,0.5}
\definecolor{lightgreen}{rgb}{0.5,1,0.5}
\definecolor{lightblue}{rgb}{0.5,0.5,1}
\definecolor{lightcyan}{rgb}{0.5,0.75,0.75}
\definecolor{lightmagenta}{rgb}{0.75,0.5,0.75}
\definecolor{customgreen}{rgb}{0.494,1,0.502}
\usepackage{url}
\usepackage{graphicx}
\usepackage{booktabs}
\usepackage{amsmath}
\usepackage{amssymb}
\usepackage{amstext}
\usepackage{amsfonts}
\usepackage{amsthm}
\usepackage{hyperref}
\usepackage{appendix}
\usepackage{physics}
\usepackage[utf8]{inputenc}
\usepackage{soul}
\usepackage{hhline}
\usepackage{comment}
\usepackage{color}
\usepackage{bm}
\usepackage[normalem]{ulem}
\usepackage{aas_macros}

\newcommand\eea{\end{eqnarray}}
\newcommand\bea{\begin{eqnarray}}

\newcommand{\be}{\begin{equation}}
\newcommand{\ee}{\end{equation}}
\newcommand{\nl}{\nonumber \\}
\newcommand{\Eq}[1]{Eq.~\ref{eq:#1}}
\newcommand{\Eqs}[2]{Eqs.~\ref{eq:#1} and \ref{eq:#2}}

\newcommand{\DM}{{_\text{DM}}}
\newcommand{\cm}{\text{cm}}
\newcommand{\GeV}{\text{GeV}}
\newcommand{\yr}{\text{yr}}
\newcommand{\hr}{\text{hr}}
\newcommand{\kpc}{\text{kpc}}
\newcommand{\AU}{\text{AU}}
\newcommand{\Hz}{\text{Hz}}
\newcommand{\GHz}{\text{GHz}}
\newcommand{\MHz}{\text{MHz}}
\newcommand{\kHz}{\text{kHz}}
\newcommand{\Fig}[1]{Fig.~\ref{fig:#1}}
 
\newcommand{\Sec}[1]{Sec.~\ref{sec:#1}}
 
\newcommand{\App}[1]{Appendix~\ref{app:#1}}

\begin{document}

\preprint{FERMILAB-PUB-26-0005-SQMS-T}
\preprint{LITP-26-02}

\title{Narrowing Down Sources of High-Frequency Gravitational Waves}

\author{Asher Berlin}
\email{aberlin@fnal.gov}
\affiliation{Theoretical Physics Division, Fermi National Accelerator Laboratory, Batavia, IL 60510, USA}
\affiliation{Superconducting Quantum Materials and Systems Center (SQMS), Fermi National Accelerator Laboratory, Batavia, IL 60510, USA}

\author{Dawid Brzeminski}
\email{dawid@umich.edu}
\affiliation{Leinweber Institute for Theoretical Physics, Department of Physics, University of Michigan, Ann Arbor, MI 48109, USA}

\author{Erwin H.~Tanin}
\email{ehtanin@stanford.edu}
\affiliation{Leinweber Institute for Theoretical Physics at Stanford, Department of Physics, Stanford University, Stanford, California 94305, USA}

\begin{abstract}
Detecting gravitational waves above 100 kHz would constitute a major discovery, as any observable signal would have to arise from new physics within the late universe. Although many technologies have been identified to explore this high-frequency regime, the known landscape of promising sources remains extremely sparse. In this work, we aim to rectify this issue by  providing model-independent arguments that highlight the most interesting parts of theory space, while remaining agnostic of the specific signal mechanism. 
For example, energy-conservation implies that gravitational waves detectable by future experiments well above a MHz would most likely have to originate from within the Solar System. Based on these arguments, we also constrain the physical properties of such sources.
\end{abstract}

\maketitle
%\tableofcontents

%%%%%%%%%%%%%%%%%%%%%%%%%%%%%%%%%%%%%%%%%%%%%%%%%
\section{Introduction}
%%%%%%%%%%%%%%%%%%%%%%%%%%%%%%%%%%%%%%%%%%%%%%%%%

Gravitational waves (GWs) offer a unique avenue to study the universe. Since LIGO's first detection of astrophysical GWs~\cite{LIGOScientific:2016aoc}, there has been a surge of interest to look for signals of cosmological origin, such as those arising from first-order phase transitions, topological defects, or cosmic reheating. Since the wavelength of these signals is limited by causality to be smaller than the comoving horizon during the time of production, GWs produced sufficiently early during the first moments of the universe are guaranteed to have a frequency well above the $1 \ \Hz - 10 \ \kHz$ range probed by the existing ground-based interferometers LIGO, Virgo, and KAGRA. It is therefore of no surprise that there has been a growing interest in identifying both: 1) new technologies to search for high-frequency GWs (HFGWs), and 2) theory-targets to motivate such searches~\cite{Aggarwal:2020olq,Aggarwal:2025noe, Gatti:2024mde,Cruise:2012zz}. 

Many promising technologies have been identified, such as those already in use by experiments searching for ultralight wavelike dark matter (DM). On the contrary, the theory landscape of detectable high-frequency sources has remained extremely sparse.  
This difficulty is best exemplified by the fact that the most optimistic futuristic setups have projected sensitivities to primordial signals many orders of magnitude weaker than cosmological bounds on the expansion rate~\cite{TitoDAgnolo:2024res,Planck:2018vyg,Figueroa:2019paj}. 
Indeed, the future detection of primordial HFGWs by any currently proposed technology  would imply the existence of gravitational radiation that overcloses the universe by many orders of magnitude.

However, HFGWs that arise \emph{locally} within our galaxy are not subject to these tight cosmological bounds. Analogous to the fact that the Milky Way is host to a $\sim 10^5$ DM overdensity, Galactic sources could give rise to a substantial overdensity of gravitational radiation whose observation has the potential to probe parts of our universe that are inaccessible through other means, such as the opaque interiors of dense astrophysical objects or regions of high DM concentration.
It therefore behooves us to remain open-minded concerning the possible signals that HFGW searches might reveal. 

Our goal in this work is to identify detectable and viable theory-targets for HFGWs. Rather than proposing specific sources, we start from the most general of considerations,  and progressively limit the viable theory space, attempting to remain as model-agnostic as possible. For instance, we show that regardless of the specific production mechanism, the large energy flow associated with  the propagation of HFGWs can lead to measurable consequences. This can be used to place significant limits on Galactic sources. Furthermore, extensive orbital measurements of large astrophysical bodies can limit the landscape of nearby sources, even if they interact only gravitationally. We will show that such model-independent considerations already significantly narrow down the likely theory-space for HFGWs.

The rest of this paper is organized as follows. In \Sec{SNR}, we outline the requirements for a source to be detectable by a future HFGW experiment.  In \Sec{EnergyArguments}, we show that bounds on energy-loss from the Earth, Solar System, and Milky Way significantly constrain the detectable parameter space of HFGW sources, solely based on their gravitational interactions. In \Sec{SourceProperties}, we specify a simple parametrization for such sources, and show that the most detectable ones correspond to  dark composite objects residing near the center of the Earth or Sun. We then discuss the plausibility of this scenario for sources in the form of spinning prolate spheroids. We then conclude in \Sec{conclusion}.

%%%%%%%%%%%%%%%%%%%%%%%%%%%%%%%%%%%%%%%%%%%%%%%%%
\section{Criteria for Detectability}
\label{sec:SNR}
%%%%%%%%%%%%%%%%%%%%%%%%%%%%%%%%%%%%%%%%%%%%%%%%%

We begin here by specifying our criteria for the detectability of HFGWs, such as requirements on the signal-to-noise ratio (SNR) and the signal occurrence rate $\Gamma_\text{event}$. In particular, we require that $\text{SNR} \gtrsim 1$, and that repeated signals of the same type occur sufficiently frequently that there is a reasonable chance for at least one to occur during a reasonable observation time.

A general way to characterize the SNR squared in a matched filter analysis is given by~\cite{Moore:2014lga}
\be
\label{eq:SNR1}
 \text{SNR}^2 = 4 \, \int\limits_0^\infty df' ~ \frac{|\tilde{h}(f')|^2}{S_n(f')}
~,
\ee
where $\tilde{h} (f)$ is the Fourier transform of the GW strain and $S_n(f)$ is the power spectral density of the strain-equivalent noise at frequency $f$. $S_n(f)$ is strongly detector-dependent, but as a benchmark value, we note that future projections  quote values typically not much smaller than $S_n (f) \sim 10^{-40} \ \Hz^{-1}$ for $f \sim 100 \ \kHz - 1 \ \GHz$~\cite{Aggarwal:2025noe}. An approximate threshold for detectability is then defined as $\text{SNR} \gtrsim 1$. Note that the integrand of \Eq{SNR1} is often written in terms of signal and noise outputs after being filtered by the transfer function of the experimental device, which converts the GW strain tensor into a scalar output. In this sense, ``$h$" in \Eq{SNR1} schematically refers to the typical value of the non-zero entries in the GW strain tensor in transverse-traceless gauge, assuming an $\mathcal{O}(1)$ dimensionless coupling to the detector.

As a first example, consider a monochromatic source with waveform $h(t)\approx h_0 \, \cos(2\pi f t)$. In this case, we can use Parseval's theorem to rewrite \Eq{SNR1} in the time domain as
\begin{align}
\label{eq:snrmonochromatic}
\text{SNR}^2 &\approx \frac{2}{S_n(f)} \, \int_0^{t_\text{obs}} \hspace{-0.18cm} dt'\,h^2(t')
\sim \frac{h_0^2\, \min{(t_\text{obs}, t_\text{sig})}}{S_n(f)}
~,
\end{align}
where $t_\text{obs}$ is the observation time and $t_\text{sig}$ is the duration of the signal. Throughout this work, we fix $t_\text{obs} = 1 \ \yr$. For burst-like sources (i.e, $t_\text{sig} < t_\text{obs}$) with frequency $f$, it is convenient to define the root-sum-squared amplitude~\cite{Moore:2014lga}
\be
h_{\rm rss}^2 = \int_0^{t_\text{sig}} dt \, h^2(t)  
    ~,
\ee
such that the SNR is given by
\be
\label{eq:snr burst}
\text{SNR}^2 \approx \frac{2 \, h_{\rm rss}^2}{S_n(f)}
~.
\ee

The strongest possible HFGW signal emerging from a system corresponds to releasing the largest possible amount of energy in a single burst. While it is not impossible that such a cataclysmic event might take place during the observing time of a HFGW experiment, such a scenario might be exceedingly unlikely. In this work, we are primarily interested in signals that are detectable not only in principle but also with a reasonable chance. To ensure that, we impose the condition that the event rate $\Gamma_\text{event}$ satisfies
\be
\label{eq:minrate}
\Gamma_\text{event}\gtrsim\text{min} \big( t_\text{obs}^{-1} \, , \,  t_\text{sig}^{-1} \big)
~.
\ee
For a burst-like signal with $t_\text{sig} \lesssim t_\text{obs}$, this corresponds to multiple events occurring within an observation time. In the other limit, for a long continuous signal $t_\text{sig} \gtrsim t_\text{obs}$, we demand that there are always multiple detectable  sources actively emitting within the duration time of the signal. Therefore, for either burst-like or continuous sources, we do not consider signals that have periods of undetectability for longer than the observation time.

%%%%%%%%%%%%%%%%%%%%%%%%%%%%%%%%%%%%%%%%%%%%%%%%%
\section{Energetics}
\label{sec:EnergyArguments}
%%%%%%%%%%%%%%%%%%%%%%%%%%%%%%%%%%%%%%%%%%%%%%%%%

The goal of this section is to determine the allowed theory-space of HFGWs on the grounds of energy conservation. Here, we remain agnostic about the emission mechanism, which is discussed in more detail in \Sec{SourceProperties}. We instead start by explaining how energetic arguments can place upper limits on the strength of detectable HFGW signals, regardless of their origin.

In particular, we will utilize the fact that a GW with characteristic frequency $f$ carries an energy density $\rho_\text{GW} \sim (h f)^2 / G$. 
This implies that at Earth, the local energy density in GWs time-averaged over $t_\text{obs}$ can be related to the SNR measure of the previous section by
\begin{align}
(\overline{\rho}_{\text{GW}})_\oplus &\sim \text{SNR}^2 \, \frac{f^2 \, S_n (f)}{G \, t_\text{obs}}
\gtrsim 10^2 \ \text{meV}^4   \bigg( \frac{f}{\MHz} \bigg)^2  
\, ,
\end{align}
where we fixed $\text{SNR} \gtrsim 1$, $S_n (f) \gtrsim 10^{-40} \ \Hz^{-1}$, and $t_\text{obs} = 1 \ \yr$ in the inequality. This will serve as a quick reference for estimating the local energy density required for detectable GWs.

As mentioned above, strong cosmological bounds on primordial GW backgrounds are easily evaded by considering local sources. 
In our investigation of local sources, we divide our surroundings into three systems, each with an increasingly smaller size and energy reservoir: the Milky Way, Solar System, and Earth. Gravitational probes allow us to infer the total mass inside of a particular region. Since GWs cannot be contained, a system enclosed by a surface $S$ of area $A_S$ that emits GWs loses mass/energy at a rate of $\dot{M} \sim - (\rho_\text{GW})_S \, A_S$.
% %
Sufficiently strong emission in GWs can lead to a noticeably reduced gravitational potential, which is indirectly probed by measurements of anomalous mass loss.

As discussed in \App{local dark masses}, there exists decades worth of data concerning the motion of astrophysical bodies within our galaxy. 
For instance, changes to the total Galactic mass can alter stellar radial velocities, which are constrained by kinematic data or by the lifetimes of various stellar systems~\cite{Coccia:2004gw}. 
At smaller scales, ephemerides of inner planets~\cite{Pitjeva:2013xxa}, helioseismology~\cite{Raffelt:1996wa}, and lunar-ranging observations~\cite{Muller:2007zzb} constrain the mass loss rate of the Solar System (i.e., Sun) and Earth. 
Such measurements are excellent probes of local HFGW sources, since they are sensitive to the average energy emitted within a given observation period. 

Existing limits on the time-averaged mass loss for the three systems discussed above are summarized as
\begin{align}
\label{eq:MdotUB}
\frac{|\langle \Dot{M}_{\rm MW} \rangle |}{M_{\rm MW}} &\lesssim 10^{-11} \ \yr^{-1} \qquad (\text{Milky Way~\cite{Coccia:2004gw}})
\nl
\frac{| \langle \Dot{M}_\odot \rangle |}{M_\odot} &\lesssim  10^{-13} \ \yr^{-1} \qquad (\text{\text{Solar System}~\cite{Pitjeva:2013xxa,Raffelt:1996wa}}) 
\nl
\frac{|\langle \Dot{M}_\oplus \rangle |}{M_\oplus} &\lesssim 7 \times 10^{-13} ~ \yr^{-1} \qquad (\text{Earth~~\cite{Muller:2007zzb}})
~,
\end{align}
where $M_{\rm MW}$, $M_\odot$, and $M_\oplus$ are the total mass of the Milky Way, the inner Solar System (within $2\ \AU$ of the Sun), and the Earth (within $60 \ R_\oplus$, corresponding to the Moon's orbital radius), respectively. Note that these limits allow for the possibility of enclosed dark masses. In what follows, we consider the implications of \Eq{MdotUB} on continuous ($t_\text{sig}\gtrsim t_\text{obs}$) or burst-like ($t_\text{sig}\lesssim t_\text{obs}$) sources.

%%%%%%%%%%%%%%%%%%%%%%%%%%%%%%%%%%%%%%%%%%%%%%%%%
\subsection{Continuous Signals}
%%%%%%%%%%%%%%%%%%%%%%%%%%%%%%%%%%%%%%%%%%%%%%%%%

For a continuous signal, the signal lasts longer than the observation time,  $t_\text{sig} \gtrsim t_\text{obs}$. Assuming isotropic power emission, the energy density in GWs evaluated at Earth's surface is related to the rate of mass loss of the source a distance $d$ away by $(\rho_\text{GW})_\oplus \approx |\langle \Dot{M} \rangle| / (4 \pi \, d^2)$. Also using that $(\rho_\text{GW})_\oplus \sim (h f)^2 / G$, the amplitude of a continuous signal can be expressed in terms of the average rate of mass loss by
\be
\label{eq:GenCont}
h_0 \sim \sqrt{G \, | \langle \dot{M} \rangle |} \, / \, (f  d)
~.
\ee
This allows constraints on $| \langle \dot{M} \rangle |$ for the various systems of interest in \Eq{MdotUB} to be recast as upper bounds on the GW strain.

%%%%%%%%%%%%%%%%%%%%%%%%%%%%%%%%%%%%%%%%%%%%%%%%%
\subsection{Burst Signals}
%%%%%%%%%%%%%%%%%%%%%%%%%%%%%%%%%%%%%%%%%%%%%%%%%

For a burst, the duration of the signal is shorter than the observation time, $t_\text{sig} \lesssim t_\text{obs}$. In this case, we use that the  average mass loss is related to mass lost in a single burst by $\langle \dot{M} \rangle \sim \dot{M}_\text{burst} \, t_\text{sig} \, \Gamma_\text{event}$, and that the energy density in GWs during the burst is once again given by $(\rho_\text{GW})_\oplus \sim |\dot{M}_\text{burst}| / d^2 \sim (h f)^2 /G$. This allows the root-sum-squared amplitude to be rewritten as
\be
\label{eq:GenBurst}
h_\text{rss} \sim \sqrt{G \, |\langle \dot{M} \rangle| \, \Gamma_\text{event}^{-1}} \, \ / \, (f d)
~.
\ee
As mentioned in the previous section, such a signal needs to be recurrent to be detectable. Thus, in addition to imposing \Eq{MdotUB} on $| \langle \dot{M} \rangle |$, we restrict $\Gamma_\text{event} \gtrsim 1 \ \yr^{-1}$, which yields an upper bound on $h_\text{rss}$ for burst-like events. 

While the limit on the Milky Way mass loss can be directly applied to this expression, there is a caveat when interpreting the limits derived from dynamics within the Solar System. In particular, orbital data has been analyzed under the assumption of continuous mass loss~\cite{Pitjeva:2013xxa, Muller:2007zzb}, which directly constrains GW sources that are active throughout the entire period of astronomical observations. We expect that these limits do not change by more than an $\mathcal{O}(1)$ number if the GW emission occurs in several periodic bursts of comparable magnitude. Therefore, for simplicity, we apply these limits to burst-like signals as well and leave a dedicated analysis to future work.

\begin{figure*}
    \centering    
    \includegraphics[width=\columnwidth]{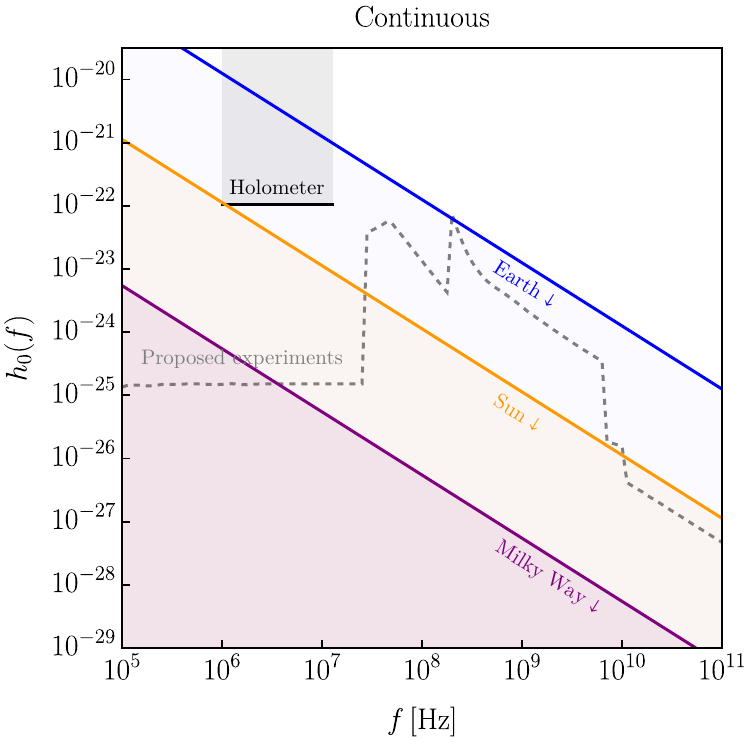}
    \includegraphics[width=\columnwidth]{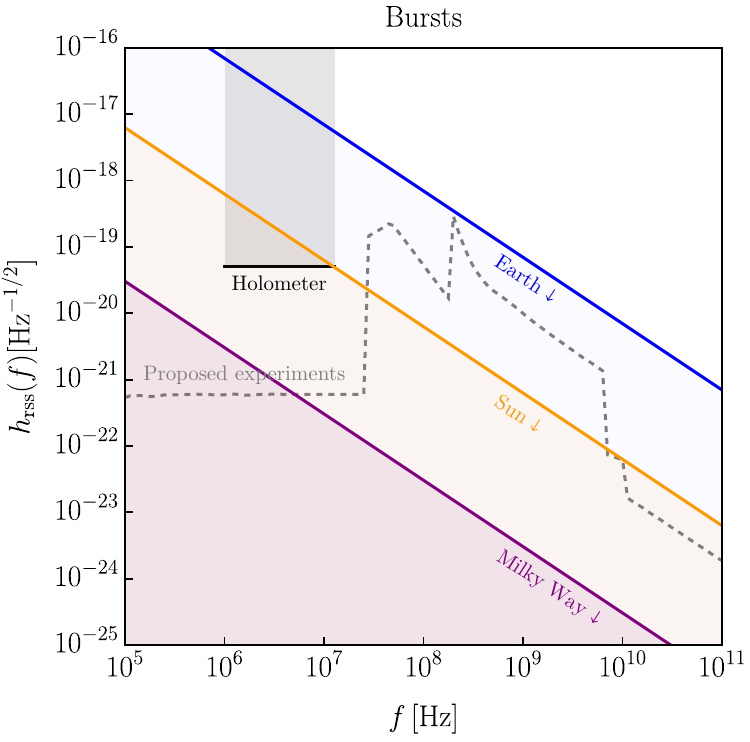}
    \caption{Viable parameter space and experimental sensitivity to the strain amplitude $h_0$ (left panel) or root-sum-squared amplitude $h_\text{rss}$ (right panel) for a continuous (left panel) or burst-like (right panel) signal of frequency $f$. The colored shaded regions indicate parameters consistent with the mass-loss constraints of \Eq{MdotUB}, for sources localized within the Earth (blue), a Sun-Earth distance (orange), or within the Milky Way (purple). For burst-like signals in the right panel, we also demand that there is on-average at least one event per observation time $t_\text{obs} = 1 \ \yr$. For comparison, we overlay limits from the Holometer experiment~\cite{Holometer:2016qoh} (shaded black) as well as projected sensitivities of various proposed detectors~\cite{Domcke:2024mfu,Schnabel:2024hem,Domcke:2024eti,Domcke:2023bat,DeMiguel:2023nmz,Ringwald:2020ist} (collectively shown as the dashed gray line). 
   } 
    \label{fig:cw}
\end{figure*}

%%%%%%%%%%%%%%%%%%%%%%%%%%%%%%%%%%%%%%%%%%%%%%%%%
\subsection{Discussion}
%%%%%%%%%%%%%%%%%%%%%%%%%%%%%%%%%%%%%%%%%%%%%%%%%

It is worthwhile to compare our model-independent considerations to specific examples. For instance, if DM possesses a non-minimal gravitational coupling, then its decay or annihilation in the Galaxy could potentially give rise to continuous HFGWs with frequency comparable to the DM mass. In this case, the local energy density in GWs can be written in terms of the DM $D$-factor (\mbox{$D_\DM \sim 10^{23} \ \GeV / \cm^2$}) or $J$-factor (\mbox{$J_\DM \sim 10^{24} \ \GeV^2 / \cm^5$})~\cite{Safdi:2022xkm}, 
\begin{align}
(\rho_\text{GW})_\oplus &\sim D_\DM \, \Gamma_\DM 
\qquad \, \text{(DM decay)}
\nl
(\rho_\text{GW})_\oplus &\sim J_\DM \, \sigma v_\DM / f
\quad \text{(DM annihilation)}
~,
\end{align}
where $\Gamma_\DM$ and $\sigma v_\DM$ are the DM decay and annihilation rates to gravitons, respectively. These constitute specific mechanisms to transfer Galactic mass into HFGWs, and thus $\Gamma_\DM$ and $\sigma v_\DM$  are constrained by the same bound on $|\langle\dot{M}_\text{MW}\rangle|$ as in the first line of \Eq{MdotUB}. 

As an example of a burst, we can consider the possibility that a substantial fraction of the energy released by a supernova explosion is carried by HFGWs, e.g., due to additional dark sector dynamics within the hot environment of the proto-neutron star. Using that this additional power in HFGWs must be smaller than the observed neutrino luminosity of SN1987A (\mbox{$\sim 10^{52} \ \text{erg} / \text{s}$})~\cite{Raffelt:1996wa}, we find that the upper bound on $h_\text{rss}$ is roughly two orders of magnitude stronger than that imposed by the Milky Way constraint in \Eq{MdotUB}. These examples serve to illustrate that our model-independent constraints are conservative and apply broadly to local sources of any type.

We now present our results for continuous and burst-like signals in the left and right panel of \Fig{cw}, respectively. Regions of parameter space that are generated by local sources with an event rate of $\Gamma_\text{event} \gtrsim 1 \ \yr^{-1}$ and obey the model-independent bounds in \Eq{MdotUB} are shown by the regions labeled ``Milky Way," ``Sun," and ``Earth," where we have used \Eqs{GenCont}{GenBurst} with $d = 10 \ \kpc$, $d = 1 \ \AU$, and $d = R_\oplus$, respectively. 

Also shown in \Fig{cw} are a collection of projections for future experimental proposals. To determine future sensitivities, we take the projections of $S_n(f)$ from Fig.~1 of Ref.~\cite{Aggarwal:2025noe} (multiplied by a factor of two to account for our conventions in \Eq{SNR1}), and use \Eqs{snrmonochromatic}{snr burst} with $t_\text{obs} = 1 \ \yr$ and $\text{SNR} = 1$. Instead, the sensitivity for the Holometer experiment corresponds to an existing limit, for which we  use $t_\text{obs} = 130 \ \hr$~\cite{Holometer:2016qoh}. Note that these projections assume broadband readout, such that $t_\text{obs}$ corresponds to the total experimental run time. 

From \Fig{cw}, we see that for $f \lesssim 10 \  \MHz$, future projections are sensitive to local sources originating out to distances as far as the Galactic Center, whereas for $f \gtrsim 1 \ \GHz$, detectable sources are required to arise within the Solar System. For burst-like signals, it is possible that viable sources generate GWs with larger strain than those highlighted here. However, this is only possible by relaxing our requirement on $\Gamma_\text{event} \gtrsim 1 \ \yr^{-1}$ or by chance a source is much closer than its typical distance, such that a successful detection is increasingly unlikely within a reasonable observation time.

In the next section, we explore how these considerations can be used to constrain specific source properties (such as mass, size, and distance).

\section{Possible Sources}
\label{sec:SourceProperties}

\subsection{$(M,L,f,d)$ Parametrization}
\label{sec:MLFd}

Here, we specify the nature of detectable HFGW sources to a greater degree. We will focus on the most optimistic case where the signal rate saturates the lower bound of \Eq{minrate}, which implies an $\mathcal{O}(1)$ probability for a signal occurring within an observation time of $t_\text{obs} = 1 \ \yr$. For simplicity, we also assume  that only a single source is dominant at a given time and that it obeys the rules of special and general relativity. Specifically, any of its parts cannot be moving faster than the speed of light, and if the source is inside its own event horizon then its properties must saturate to those of a black hole. 

Moreover, we assume that the sources are isolated, and hence have exhaustible kinetic energy.\footnote{It is logically possible that the internal motion of the source is powered by a mechanism that converts rest mass into kinetic energy or that the source acquires energy from its surrounding. We do not consider such cases here.}  Thus, a stronger strain implies a shorter signal, since the power emitted in GWs drains the source's kinetic energy. Furthermore, as discussed in the previous section, GW emission decreases the source's gravitating mass, which for local signals is constrained by the measured orbits of planets, asteroids, and satellites, as well as long-term stability arguments. We will show that these model-independent considerations place significant constraints on the sources of \emph{detectable} HFGWs.

Localized sources undergoing non-spherical periodic internal motion with time-dependent quadrupole moments can be parametrized with the parameter set $(M,L,f,d)$. Here, $M$ is a fraction of the source mass that moves coherently over a length scale $L$ and harmonically with frequency $f$, and $d$ is the distance to the source. Note that $L$ parametrizes the aspherical size of the source; a near-spherical source could be large overall while having a small $L$. In order to maximize the strength of these signals, we will assume that the asphericity of the source is $\mathcal{O}(1)$. 

In the limit that the source is non-relativistic ($v\sim Lf\ll 1$), the GW wavelength is large compared to the source size ($f^{-1}\gg L$). If the wavelength is also small compared to the source distance ($f^{-1}\ll d$) (i.e., the radiation zone), the GW strain is given by
\be
\label{eq:hquad}
h\sim \frac{G\ddot{Q}}{d} \sim \frac{GM (f L)^2}{d}\sim \frac{GMv^2}{d} 
~,
\ee
where $Q \sim M L^2$ is the mass quadrupole moment. Causality restricts the internal velocity of the source to be \mbox{$v\sim Lf\lesssim 1$}.
The Schwarzschild radius $r_s = 2 G M$ sets the minimum possible size of the moving source mass \mbox{$L\gtrsim r_s$}. These restrictions can be reexpressed as a limitation on the frequency $f \lesssim 100 \ \kHz \times (M / M_\odot )$ as well as the size of the source 
\be
\label{eq:Lrange}
3\ \text{km} \times \left(\frac{M}{M_\odot}\right)\lesssim L\lesssim 300\ \text{km} \times \left(\frac{f}{\kHz}\right)^{-1} 
~.
\ee

We optimistically assume that the source loses its internal energy dominantly into the production of GWs, although below we show that there can be additional inevitable sources of dissipation (e.g., due to the emission of sound-waves). A strong strain corresponds to a large instantaneous luminosity in GWs $\mathcal{L}_\text{GW}\sim G\dddot{Q}^2\sim   GM^2L^4f^6$, which can deplete the kinetic energy of the source $\sim M L^2 f^2$ relatively quickly. This sets an upper bound on the duration of the GW signal $t_\text{sig}\lesssim M (L f)^2/\mathcal{L}_\text{GW}\sim (GML^2f^4)^{-1}$, corresponding to 
\be
\label{eq:tsig}
t_\text{sig}\lesssim 10 \ \text{ms} \times \left(\frac{M}{M_\odot}\right)^{-1}\left(\frac{L}{\text{m}}\right)^{-2}\left(\frac{f}{\MHz}\right)^{-4}
~.
\ee

As discussed in \Sec{EnergyArguments}, there are also limits on the total energy-loss rate of the Milky Way, Sun, and Earth. In particular, the bounds in \Eq{MdotUB} correspond to a maximum allowed GW luminosity time-averaged over a year, $\bar{\mathcal{L}}_\text{GW} \lesssim \bar{\mathcal{L}}_\text{GW}^\text{max}$, where
\begin{align}
\bar{\mathcal{L}}_\text{GW}^\text{max} \equiv
\begin{cases}
        1\times 10^{48}\ \text{erg/s} &\text{(Milky Way)}\\
        6\times 10^{33}\ \text{erg/s} &\text{(Solar System)} \\
        1\times 10^{29}\ \text{erg/s} &\text{(Earth)}
        ~.
    \end{cases}
    \label{eq:LGWbarlimits}
\end{align}
Note that the above bound for the Solar System is weaker than that typically used when considering the production of new feebly-coupled particles in the Sun~\cite{Raffelt:1996wa}. This difference is due to the fact that we are not necessarily interested in processes that carry thermal energy away from the Solar interior. For instance, GWs may arise from the mass-energy of exotic dark objects trapped within the Solar System. Thus, \Eq{LGWbarlimits} is more conservative. 

For an event rate $\Gamma_\text{event}$ satisfying \Eq{minrate}, the average luminosity in detectable GWs is related to the instantaneous luminosity by
\be
\label{eq:LGWbar}
\bar{\mathcal{L}}_\text{GW} \sim  \mathcal{L}_\text{GW} \, t_\text{sig} \, \Gamma_\text{event}
\gtrsim
\mathcal{L}_\text{GW} \, \frac{\text{min}(t_\text{sig},t_\text{obs})}{t_\text{obs}}
~,
\ee
where we fix the observation time to \mbox{$t_\text{obs} = 1 \ \yr$}. To be detectable, we also require the $\text{SNR} \gtrsim 1$ in \Eq{snrmonochromatic}, which along with \Eq{hquad} gives
\begin{align}
    M &\gtrsim M_\odot  \times \bigg( \frac{f}{\MHz} \bigg)^{-2}  \bigg( \frac{L}{\text{m}} \bigg)^{-2}  \bigg( \frac{d}{10 \ \text{pc}} \bigg) 
    \nl
    &\bigg( \frac{\min(t_\text{sig}, t_\text{obs})}{10 \ \text{ms}} \bigg)^{-1/2}  \bigg( \frac{S_n}{10^{-40} \ \Hz^{-1}} \bigg)^{1/2}
    ~.
    \label{eq:detectable}
\end{align}

\begin{figure*}[t!]
    \centering
    \includegraphics[width=0.495\linewidth]{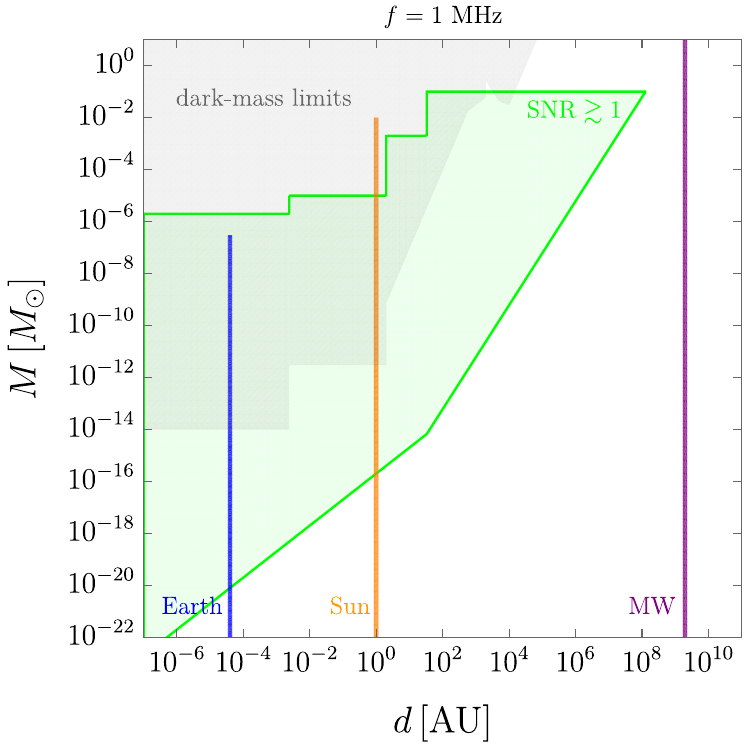}
    \includegraphics[width=0.495\linewidth]{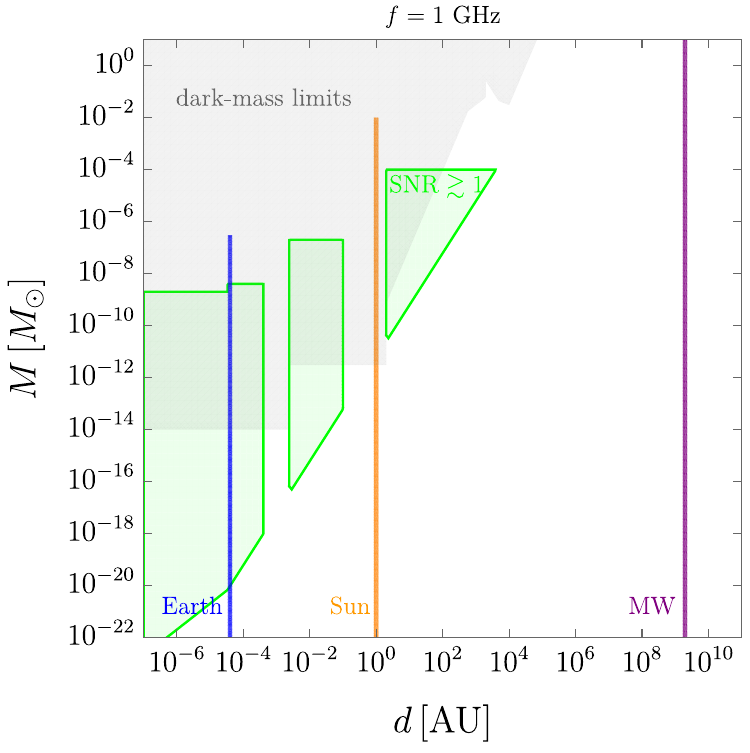}
    \caption{Allowed values of the source mass $M$ and distance to the source $d$ for a gravitational wave signal frequency of $f = 1  \ \MHz$ (left panel) or $f = 1 \ \GHz$ (right panel). For each choice of $M$ and $d$, the size $L$ of the source is allowed to float in the range consistent with the restrictions of \Eq{Lrange}. In the green region, sources satisfying the self-consistency and energy-loss constraints of \Sec{MLFd} are detectable by an experiment with an observation time of $t_\text{obs} = 1 \ \yr$ and strain-equivalent noise spectral density $S_n = 10^{-40} \ \Hz^{-1}$. 
    In the shaded gray regions, we show a collection of bounds on the static effect of exotic dark masses bound to the Solar System that extend outside of the Sun or Earth. 
    These include limits from planets, asteroids, spacecraft, Voyager ranging to Uranus, as well as Moon and Earth-bound artificial satellites. The blue, orange, and purple regions show the allowed range of masses for a source bound to the center of the Earth, Sun, and Milky Way, respectively. 
}
    \label{fig:Mvsd}
\end{figure*}

\subsection{The Space of Detectable High-Frequency Gravitational Wave Sources}
In \Fig{Mvsd}, we show the parameter space spanned by the source mass $M$ and distance $d$, fixing the signal frequency to $f=1\ \MHz$ (left panel) or $f = 1 \ \GHz$ (right panel). The size $L$ of the source is allowed to float in the range consistent with the restrictions of \Eq{Lrange}.

In the shaded green region, we highlight parameter space that is not ruled out by the energy-loss bounds of the previous section and that is detectable by an experiment with noise spectral density $S_n=10^{-40}\ \Hz^{-1}$ and operation time $t_\text{obs}=1\ \yr$. In particular, we only highlight a parameter point with fixed $M$ and $d$ if there is a possible value of $L$ that is both consistent with the limits of \Eq{LGWbarlimits} and yields $\text{SNR} \gtrsim 1$. Note that although our adopted noise-level value $S_n$ is roughly two orders of magnitude larger than the strongest projections from Ref.~\cite{Aggarwal:2025noe}, this choice is more representative of the full suite of projections across the entire $\MHz - \GHz$ frequency range.

The non-trivial features in the boundary of this green region are set by the following requirements:
\begin{itemize}
\item \textit{Detectability} ($\text{SNR}\gtrsim 1$): The minimum source mass in \Eq{detectable} scales as $M_{\rm min}\propto d$ for $t_\text{sig}\gtrsim t_\text{obs}$ and $M_{\rm min}\propto d^2$ for $t_\text{sig}\lesssim t_\text{obs}$. 
\item \textit{Causality and BH Saturation}: Requiring the maximum size $L$ in \Eq{Lrange} be greater than the minimum size leads to an upper limit on the viable source mass, $M \lesssim (G  f)^{-1} \sim 0.1 \ M_\odot \times (\MHz / f)$.   
\item \textit{GW Luminosity Bound}: 
For certain parameters, a stronger upper limit on the mass arises from the GW luminosity bound of \Eq{LGWbarlimits}. In this case, setting $L$ to its minimum size and assuming that the signal duration is governed by energy loss into GWs $t_\text{sig} \sim M (L f)^2 / \mathcal{L}_\text{GW}$ as in \Eq{tsig}, we find 
\begin{align}
\label{eq:MUB}
M\lesssim 
\begin{cases}
(\bar{\mathcal{L}}_\text{GW}^\text{max} \, t_\text{obs} / G^2 \, f^2)^{1/3} &(t_\text{sig}\lesssim t_\text{obs})\\
(\bar{\mathcal{L}}_\text{GW}^\text{max}/G^5 \, f^6)^{1/6} & (t_\text{sig}\gtrsim t_\text{obs})
~.
\end{cases} 
\end{align}
This bound applies to the case where gravitational radiation is dominantly responsible for setting the signal duration $t_\text{sig}$, as is the case for, e.g., merging binaries. Hence, \Eq{MUB} can be weakened if we assume that $t_\text{sig}$ is parametrically shorter than the time it takes for gravitational radiation to drain the source's kinetic energy (note, however, that shortening $t_\text{sig}$ also reduces the SNR). This can occur if, e.g., the signal originates within the interior of the Sun or Earth, in which case a substantial fraction of the source's energy is dissipated into sound waves. We will investigate this example in more detail in the next subsection. Bounds on the GW luminosity from \Eq{LGWbarlimits} also limit the detectability of a given source. For instance, \Eq{LGWbarlimits} can be expressed in terms of the SNR of \Eq{snrmonochromatic} as
\begin{align}
\text{SNR} &\lesssim 10^2 \times  \left( \frac{d}{\AU}\right)^{-1} \left(\frac{f}{\MHz} \right)^{-1} \left( \frac{t_\text{obs}}{\yr} \right)^{1/2}
\nonumber\\
& \left( \frac{S_n(f)}{10^{-40}\ \Hz^{-1}} \right)^{-1/2} 
\left( \frac{\bar{\mathcal{L}}_\text{GW}^\text{max}}{6\times 10^{33}\ \text{erg/s}} \right)^{1/2}
~,
\label{eq:LGWlimitonSNR}
\end{align}
where we used that $\mathcal{L}_\text{GW} \sim 4 \pi d^2 \, (h f)^2 / G$. 
For a fixed $f$, this sets a $d$-dependent upper bound on the SNR, regardless of $M$ and $L$. Requiring that $\text{SNR} \gtrsim 1$ then leads to an upper limit on $d$. 
\end{itemize}

\begin{figure*}[t!]
    \centering
    \includegraphics[width=0.495\linewidth]{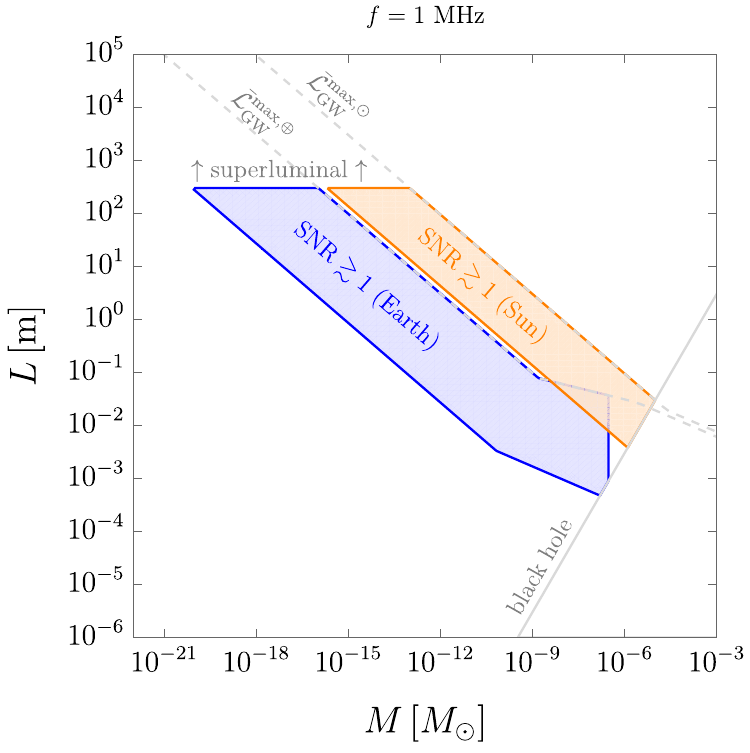}
    \includegraphics[width=0.495\linewidth]{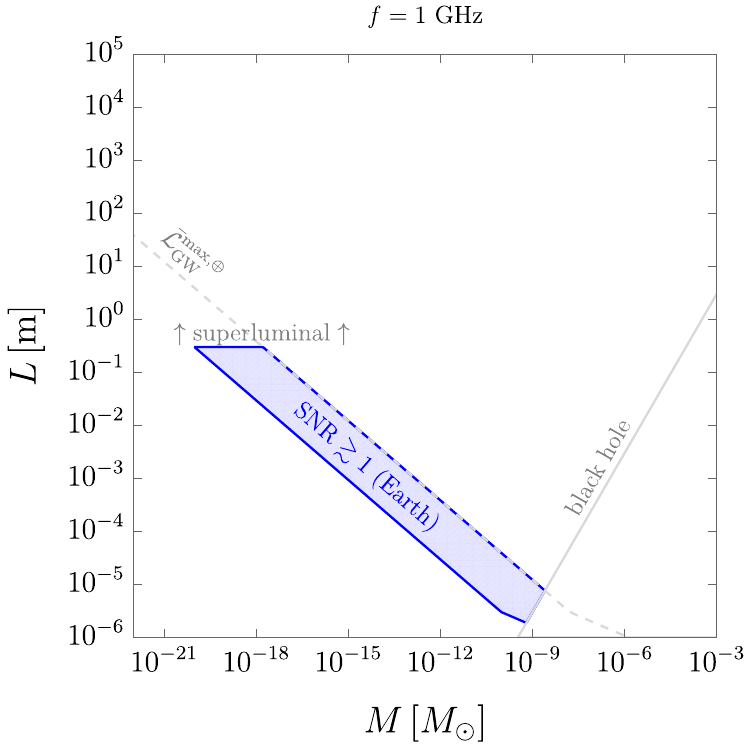}
    \caption{Allowed characteristic size $L$ and mass $M$ of detectable HFGW sources at the center of the Earth (shaded blue) or Sun (shaded orange), with frequency $f=1\ \MHz$ (left panel) or $f=1\ \GHz$ (right panel). The steeper/flatter SNR=1 lower boundaries correspond to sound-wave/GW dissipation, respectively. The source mass is limited to be $M\lesssim 10^{-1} \ M_\oplus$ inside the Earth and $M\lesssim 10^{-2} \ M_\odot$ inside the Sun. The largest possible $M$ for a given $L$ is further limited by $r_s=2GM\lesssim L$. The upper limits on $L$ are set by superluminality avoidance ($Lf\lesssim 1$), and limits on the average GW luminosity from \Eq{LGWbarlimits} (see also \Eq{LGWlimitonSNR}). For $f=1\ \GHz$, there is no viable parameter space for a detectable source located at the center of the Sun.}
    \label{fig:SSSunEarth}
\end{figure*}

We see from \Fig{Mvsd} that the majority of the detectable parameter space corresponds to light ($M \ll M_\odot$) and relatively nearby ($d \ll \kpc$) sources. In principle, a nearby signal could arise from a source following a trajectory that happens to take it close to our Sun. Although this is possible, it is highly improbable to observe such a transient GW signal from a source passing within the extended Solar System ($d \lesssim 10^3 \ \text{AU}$). This is because the likelihood to observe such an event is doubly suppressed by the requirements that the source be passing nearby on a hyperbolic trajectory while also emitting a burst of GWs during the transit. 

Nevertheless, the requirement of passing nearby can be guaranteed if the source has been on a bound orbit around the Sun for an indefinitely long time. As discussed in \App{local dark masses}, if such objects exist outside of the Sun or planetary bodies, they modify the form of the corresponding gravitational potential,
which is bounded by measurements of the orbital properties of various known objects, including artificial satellites, the Moon, asteroids, planets, and the Oort cloud. Note that unlike \Eq{LGWbarlimits}, these limits apply to orbiting bodies independent of whether or not they emit gravitational radiation, 
and, hence, are solely a function of $M$ and $d$, independent of $f$ and $L$. These bounds are displayed as the shaded gray regions in \Fig{Mvsd}.

These limits on the existence of Sun-bound dark masses can be circumvented if the dark masses are in fact trapped inside a celestial body. In this case, the static effect of the dark mass is degenerate with that of the mass of the celestial body. The Earth and the Sun are arguably the most interesting candidates to host dark masses, due to their proximity and density (which may aid in capture). Even in this case, though, there are known limits on such Earth-bound and Sun-bound masses.\footnote{Nearby stellar remnants may also serve as promising hosts of detectable HFGW sources. Although the closest known neutron star and white dwarf are located $\sim 10^{7}\ \text{AU}$ and $\sim 10^6$\ \text{AU} away, respectively, they have orders of magnitude higher densities compared to the Sun and the Earth.} We have compiled these limits in \App{local dark masses} and display the regions consistent with these limits in \Fig{Mvsd} as blue (Earth-bound) and orange (Sun-bound) regions.

\subsection{Accompanying Effects}

The discussion above implies that the center of the Earth or  Sun are optimal sites for detectable sources. In \Fig{SSSunEarth}, we show the range of viable masses $M$ and sizes $L$ of sources at the center of the Earth ($d\approx 1 \ R_\oplus$) and at the center of the Sun ($d\approx 1\ \AU$), for signal frequencies of $f=1\ \MHz$ and $f=1\ \GHz$. In this section, we discuss additional model-independent signals that such sources would automatically produce through their gravitational interactions.

To begin, consider the case with  $f=1\ \MHz$, $S_n =10^{-40}\ \Hz^{-1}$, and $t_\text{obs}=1\ \yr$. From \Fig{SSSunEarth} we see that in light of the previously discussed constraints the remaining viable parameter space for sources bound to the interior of the Sun corresponds to \mbox{$M \sim (10^{-16}-10^{-5}) \ M_\odot$} and \mbox{$L\sim (10^{-2}-10^2) \ \text{m}$}.
% %
The midpoint of these ranges ($M\sim 10^{-10} \ M_\odot$ and $L\sim 1\ \text{m}$) corresponds to an object of mass density \mbox{$\rho \sim M/L^3\sim 10^{17} \ \text{g}/\cm^3$}, orders of magnitude larger than the core density of a neutron star ($\rho_{\rm NS}\sim 10^{14}\ \text{g}/\cm^3$), internal velocity $v\sim L f \sim 10^{-3}$, and escape velocity $v_\text{esc}^2\sim GM/L\sim 10^{-7}$. 

Similarly, for Earth-bound sources, the allowed masses and sizes for $f = 1 \ \MHz$ are \mbox{$M\sim (10^{-20}-10^{-6}) \ M_\odot$} and \mbox{$L\sim (10^{-3}-10^2) \ \text{m}$}.
The approximate midpoint is $M\sim 10^{-13} \ M_\odot$ and $L\sim 0.1\ \text{m}$, which corresponds to a source with mass density  $\rho\sim 10^{17}\ \text{g}/\cm^3$, internal velocity $v\sim 10^{-4}$, and escape velocity $v_\text{esc}^{2}\sim 10^{-9}$. 

Due to their large density, such objects would presumably sink to the core of the Sun or the Earth, even if they interact only gravitationally with SM particles (see \Eq{DFtime} of \App{interiors}). Furthermore, the strong gravitational fields near these objects could stir the baryons around them, driving MHz sound waves through their time-dependent gravitational potentials $\Phi(t)$. We discuss this in more detail below.

\subsubsection{Sound Waves}

The time-dependent gravitational potential of a GW source trapped within the Sun or Earth can produce sound waves. These sound waves might be detectable, e.g., if they can propagate to Earth's surface without being completely attenuated. Regardless, even if these sound waves are not detectable, their emission contributes to the shortening of the signal duration $t_\text{sig}$. An object of mass $M$ moving with speed $v$ in the core of the Sun or Earth loses energy via gravitational excitation of sound waves at a rate of 
\begin{align}
\label{eq:soundwaves}
\mathcal{L}_{\rm SW}^{\odot,\oplus} &\sim 
\frac{4\pi G^2 M^2}{v} \, \text{min}\bigg(10 \, , \, \frac{v^3}{3 \, (c_s^{\odot, \oplus})^{3}}\bigg) \, \rho_{\odot,\oplus}
\nonumber\\
&\sim \left(\frac{L}{\text{m}}\right)^{-1}\left(\frac{f}{\MHz}\right)^{-1}\text{min}\left(10,\frac{L^3f^3}{3 (c_s^{\odot, \oplus})^{3}}\right)\nonumber\\ 
& \times \begin{cases}
2\times 10^{43}\ \frac{\text{erg}}{\text{s}} \, \left(\frac{M}{10^{-2}M_\odot}\right)^2 & (\text{Sun})
\\
2\times 10^{32}\ \frac{\text{erg}}{\text{s}} \, \left(\frac{M}{10^{-7}M_\odot}\right)^2& (\text{Earth})
~,
\end{cases}
\end{align}
where $\rho_\odot \sim 100\ \text{g}/\cm^3$ and $\rho_\oplus \sim 10\ \text{g}/\cm^3$ are the mass density in the core of the Sun and Earth, respectively, and $c_s^{\odot}\sim 1\times 10^{-3}$ and $c_s^{\oplus} \sim 3\times 10^{-5}$ are the sound speeds~\cite{Bahcall:2000nu,2016SciA....2E0802S}. \Eq{soundwaves} was obtained by multiplying the typical dynamical friction force on the source $F_{\rm DF}\sim (4\pi G^2M^2\rho_{\odot, \oplus}/v^2)\times \text{min}\left(10,v^3/3c_s^3\right)$ with the source's typical velocity $v\sim Lf$ (see \Eq{DF} of \App{interiors}). Accounting for this additional energy loss, the duration of the GW signal is shortened to
\be
t_\text{sig}\sim \frac{ML^2f^2}{\mathcal{L}_\text{GW}+\mathcal{L_{\rm SW}}}
~.
\ee
We account for this effect in \Fig{SSSunEarth}.

While GWs simply escape the Sun or the Earth, sound waves stay trapped and eventually dissipate into thermal energy. This can lead to detectable (transient or long-term) ``global warming" of the Sun/Earth. Avoiding such scenarios can in principle lead to additional constraints.

\subsubsection{Gravitational Memory}

``Gravitational memory" is a particular perturbation to the strain  produced by the time-varying quadrupole moment associated with source-components free-streaming to infinity~\cite{Favata:2010zu}. It is characterized by a permanent change in strain $\Delta h_{ij}^{\rm after}=h_{ij}^{\rm mem}(t\rightarrow+\infty)$ 
relative to that prior to the source-components flying to infinity. Here, we are mainly interested in scenarios where the free-streaming source-components are themselves GWs, in which case the resulting memory is called ``non-linear/Christodoulou" memory~\cite{Christodoulou:1991cr,Thorne:1992sdb}.  

In many cases, it is easier to detect the changing strain during the build-up of the memory, $h_{ij}^{\rm mem}(t)$, than the permanent change $\Delta h_{ij}^{\rm after}$ at the end of the process. 
As we review in \App{memory}, the memory signal can be written as a time integral over the entire history of the HFGW source. For certain GW sources (e.g., in the case of a primordial black hole merger), the memory components of HFGW signals may have lower-frequency Fourier components that are within the sensitivity range of LIGO.  
However, in general this memory signal depends on the specific envelope of the primary GWs sourcing it, meaning that it is strongly model-dependent. Hence, to keep our discussion conservative and model-independent, we do not consider memory signals in our main analysis. In \App{memory}, we illustrate this model-dependence with some specific examples.

\subsubsection{Time-Dependent Gravitational Potential} 
Expanding the Newtonian potential $\Phi(t)$ of a source in terms of its first few multiple moments gives
\be
\Phi(t) \approx 
- \frac{GM}{d} - \frac{3G \, Q_{ij}(t) \, \hat{r}^i \, \hat{r}^j}{2 \, d^3}
~.
\ee
Hence, a time-dependent quadrupole moment $Q_{ij}(t)$ inevitably produces a time-varying non-radiative Newtonian gravitational potential $\Phi(t)$ according to the above expression. While only the time-dependent quadrupole moment of a source contributes to the far-field radiative emission of  GWs, the same source may also have time-dependent dipole or even monopole moments. The time-derivatives of the dipole and monopole moments of some sources must be zero in order to conserve mass and linear momentum, but these need not apply in general. For example, the source could be undergoing bulk motion, accretion, or mass loss. To be conservative, we consider only the quadrupole part of the time-dependent Newtonian potential, \mbox{$\Phi(t)\sim GML^2/d^3$}. Such a time-dependent and spatially non-uniform Newtonian potential $\Phi(t)$ generically gives rise to a gradient in acceleration to test masses, corresponding to a fractional change in length $L_\text{det} \, \ddot{h}_\Phi\sim |\nabla\Phi| \, L_\text{det}/d$ described by an effective strain of \mbox{$h_\Phi \sim \Phi/(f d)^2 \sim h  / (f \, d)^4$}, where \mbox{$h \sim G M v^2 / d$} is the typical GW strain of a source moving with velocity $v \sim f L$ and $L_{\rm det}$ is the characteristic size of the detector. For the cases of interest here, $f d \gg 1$ and hence $h_\Phi \ll h$.

\subsection{Plausibility Arguments}

The general considerations of the previous subsections illustrate the point that the most detectable sources of HFGWs are macroscopic dark objects bound to the Solar System or inside the Earth/Sun, and undergoing some sort of oscillatory or rotational motion. In this section, we discuss qualitatively how such sources may reasonably be realized, and provide plausibility arguments for their possible origins.

\subsubsection{Source Models}

Primordial black holes (PBHs) and black hole superradiance are some of the most widely discussed sources of HFGWs~\cite{Aggarwal:2025noe}. For these cases, the signal frequency is related to the mass and size of the source by
\begin{align*}
f &\sim \sqrt{GM / L^3} &&\text{(PBH binary inspiral)}
\\
f &\sim L^{-1}\sim (GM)^{-1} &&\text{(PBH merger, superradiance)}
~.
\end{align*}

How do we populate the rest of the $(M,L)$ parameter space? For gravitationally bound objects, such as PBH binaries, we generally have $L \sim (GM/f^2)^{1/3}$. The simplest way to generate larger sources, for a given frequency and mass, is to invoke a new long-ranged attractive dark force, corresponding to $L \sim (\alpha GM/f^2)^{1/3}$, where $\alpha \gg 1$ parametrizes the strength of the new force with respect to gravity~\cite{Bai:2024pki,Bai:2023lyf,Kaplan:2024dsn,Fedderke:2024hfy,Graham:2025gtd}. However, we note that such a long-ranged force requires a light mediator field, whose emission contributes to additional dissipative power that further limits the duration of the GW signal. On the other hand, $L\lesssim (GM/f^2)^{1/3}$ corresponds to weaker-than-gravitational binding energy. This can be achieved with a competition between gravitational attraction and a repulsive dark force that becomes stronger at smaller values of $L$.

As an existence proof of a reasonable model that can cover all of the $(M,L,f,d)$ parameter space considered here, we introduce a new class of HFGW source: a rotating non-axisymmetric dark composite. For brevity, we will refer to this type of source as a spinning (American) football.\footnote{In this work, we define a ``football" as an American-like football with quadrupole moment $Q\sim ML^2$. The \textit{real} football, of course, has a suppressed quadrupole moment $Q\ll ML^2$, but the authors have been unable to reach a consensus.} Spinning footballs emit continuous and quasi-monochromatic GWs in a way analogous to rotating deformed neutron stars~\cite{1971ApJ...166..175I}. For example, a triaxial ellipsoid of mass $M$ and semi-axes $L_{1,2}$ rotating with a frequency $f$ would emit GWs with $\left<h^2\right>=2G\mathcal{L}_\text{GW}/\pi^2f^2d^2$ and $\mathcal{L}_\text{GW}=128\pi^2 GM^2f^6(L_1^2-L_2^2)^2$~\cite{1985Ap&SS.111..335B}. Footballs can in principle have any combination of $(M,L,f,d)$ as each of these parameters can be separately varied in isolation. Furthermore, they do not automatically emit radiation other than GWs. As a result, they can maintain their signal frequency for as long as the spin-down timescale, the latter of which can be governed by GW emission (or gravitational excitation of sound waves in matter).

Given any particular source of HFGWs, there are typically additional constraints if one considers the full evolution of the source before or after entering the high-frequency regime. For instance, a PBH binary evolves in a runaway inspiral which produces a high-frequency GW burst at the end of the merger. In some cases, this can lead to a low-frequency memory signal detectable by LIGO/Virgo (but this is not always the case, as discussed in \App{memory}). 

On the other hand, for a rotating football, the signal frequency  naturally decreases over time as the rotation slows down due to GW emission, eventually entering the frequency-band of LIGO/Virgo from above. As the football spins down, the signal strain evolves as \mbox{$h\sim GM( Lf)^2/d\propto f^2$}, since $M$, $L$, and $d$ are fixed. Hence, by the time the signal frequency enters the LIGO/Virgo band $f_\text{LIGO} \sim 100 \ \Hz$, its strain is reduced by a factor of $(f_\text{LIGO}/f_\text{HF})^{2}$ compared to the signal at a much higher frequency $f_\text{HF} \gg f_\text{LIGO}$. Moreover, the timescale $\tau (f)$ for GW emission to cause the football to spin down and lower its frequency by an $e$-fold becomes longer with time according to $\tau (f) \propto f^{-4}$. As an example, for a source initially at $f_0=10\ \MHz$ with $\tau (f_0) \sim 1\ \yr$, it would take $(f_0 / 10\ \kHz)^4 \, \tau(f_0) \sim 10^{3}\ \text{Gyr}$ to reach the LIGO/Virgo band, which is far longer than the age of the universe. Hence, there exists ample parameter space where the football's late-time GW emission is completely unexplored by LIGO/Virgo. Even if the football's GW signal makes it into the LIGO/Virgo band, the signal strain it emits during this  is severely reduced. A signal with $f = 1 \ \MHz$, $t_\text{sig}\sim \yr$, and $\text{SNR} = 1$ with our benchmark detector ($S_n =10^{-40}\ \Hz^{-1}$) corresponds to $h \sim 2\times 10^{-24}$. By the time the signal frequency evolves to $f = 10 \ \kHz$, the strain is $h\sim 2\times 10^{-24} \, (10 \ \kHz/1\ \MHz)^2\sim 2\times 10^{-28}$, far below the sensitivity of LIGO/Virgo.

\subsubsection{Production Mechanisms}

Rotating footballs need to have strong binding energies to maintain their non-axisymmetric shape, as well as a mechanism to generate their angular momentum. We are not aware of a concrete scenario that can produce such states in a large abundance, but it is nonetheless a logical possibility. For instance, a violent explosion or collision between two massive objects in the dark sector may eject small but macroscopic chunks of rotating composites, which may then pass near the Earth/Sun, before getting captured. Dark binaries bound by non-gravitational long-range forces may have clearer origin stories and have been considered  in Refs.~\cite{Bai:2023lyf,Bai:2024pki,Fedderke:2024hfy,Kaplan:2024dsn,Wang:2025mea}.

\subsubsection{Trapping Mechanisms}

It is possible that capture of dark objects occurs continually and is an ongoing process. Assuming these objects arrive with the typical virial velocity in the Milky Way, $v_\text{vir}\sim 10^{-3}$, and accounting for the focusing effect of possible long-range forces, the maximum possible capture cross section by the Earth or Sun is $\sim \pi R_{\oplus, \odot}^2/v_\text{vir}^2$. 

As an extreme example, consider the possibility that $10\%$ of the Galactic DM density is composed of a particle that possesses strong interactions with baryons, so that every such particle that crosses through the Earth or Sun efficiently downscatters and becomes bound. In this case, the maximally allowed capture cross-section over the $\sim 5 \ \text{Gyr}$ age of the Solar System corresponds to a total dark mass of $\sim 10^{-9} \ M_\odot$ bound to the Earth and $\sim 10^{-5} \ M_\odot$ bound to the Sun. Alternatively, capture of dark particles may also occur during the initial formation of the Solar System. However, a challenging aspect for this scenario is that the capture mechanism needs to operate in the low-density environment of the proto-Earth/Sun.

Capturing macroscopic dark masses requires some mechanism to remove their kinetic energy relative to the Earth/Sun, which can be achieved through self-interactions~\cite{Ebadi:2025umm} or interactions with baryons~\cite{Jacobs:2014yca,Ebadi:2021cte}. 
While these same couplings generically tamper with the emission of GWs (e.g., by acting as an extra source of dissipation), it is also possible that they effectively shut off after  capture. A favorable case is if the dark objects have a small velocity dispersion relative to the Solar System prior to capture. For instance, this could happen if they form a dark disk by dissipating their orbital energy around the Galactic Center~\cite{Fan:2013yva}. It is then possible that the dark mass arrives at the Earth or Sun on a near parabolic orbit, with a near-zero kinetic plus gravitational potential energy. If so, even a tiny amount of dissipation (via, e.g., gravitational dynamical friction) is enough to bring the dark mass into an elliptical orbit that passes through the Earth or the Sun repeatedly. Another possibility is if DM particles are captured via scattering with baryons~\cite{1985ApJ...296..679P}. The captured dark mass then grows with time, until eventually it crosses a threshold whereupon it collapses under its own gravity~\cite{Goldman:1989nd,1990PhLB..238..337G}. What happens next is model-dependent, but one can generically expect some emission of HFGWs, analogous to how GWs are generated in core-collapse supernovae~\cite{2002ApJ...565..430F} (see also Ref.~\cite{Kurita:2015vga}).

\section{Discussion and Conclusion}
\label{sec:conclusion}

Over the past decade, there has been a surge of activity in identifying promising technology for the detection of HFGWs. We have aimed to complement these efforts by pinning down the possible origin of detectable signals in this high-frequency regime. In doing so, we have remained agnostic of the specific models that could generate these signals. Instead, we have used general principles, such as energy conservation, to deduce the following: 
1) HFGWs that are detectable by any currently proposed technology need to be sourced locally in the late universe. 
2) The strength of such local signals are significantly constrained by the measured orbits of various spacecraft, asteroids, and planets. This implies that it is exceedingly unlikely that a signal  above $\sim 1 \ \MHz$ will be observed originating from outside the Solar System. %Similarly, above $\sim 100 \ \MHz$, a detectable source should reside within an AU. 

We have also outlined the detectable parameter space in terms of the source's mass, characteristic size, frequency, and distance. Since such sources are necessarily massive and nearby, they are subject to strong model-independent limits, solely based on their gravitational influence.   
For instance, sources on elliptical orbits around the Sun are subject to bounds derived from ephemerides and Oort cloud measurements, which we have compiled in \App{local dark masses}. However, limits from ephemerides of inner planets are alleviated if the  source resides within the interior of the Sun or  Earth, in which case they are subject to a different set of weaker constraints. These highly general considerations allow us to narrow down the possible  parameter space for detectable HFGWs. We have also briefly explored the possibility that this model space can be realized by sources in the form of dark composite rotating prolate spheroids.

We hope that our work will provide useful guidance in charting the HFGW theory space and maximizing the discovery potential of upcoming experiments. In future work, it would be interesting to move beyond the simple order-of-magnitude estimates performed here. In particular, we did not consider sources on hyperbolic orbits,  moving ultra-relativistically, or those that are actively accreting mass, which could introduce qualitative modifications to some of the claims made here.

\section*{Acknowledgments}
We thank Michael Fedderke, Peter Graham, and Ben Lehmann for useful discussions. This material is based upon work supported by the U.S. Department of Energy, Office of Science, National Quantum Information Science Research
Centers, Superconducting Quantum Materials and Systems Center (SQMS) under contract number DE-AC02-
07CH11359. E.H.T. is supported by NSF Grant PHY-2310429, Simons Investigator Award No. 824870, the Gordon and Betty Moore Foundation Grant GBMF7946. 
D.B. is supported by
the Department of Energy under grant number DE-SC0007859.
%\clearpage
%\newpage
\appendix

\section{Limits on Local Dark masses} 
\label{app:local dark masses}
\subsection{Outer Solar System}

The Oort cloud contains $\sim 10^{12}$ comets in long-period orbits around the Sun with semimajor axes $10^3-10^5\ \AU$. Its overall size sets the following limits on dark masses $M$ in orbits around the Sun with semimajor axes $a$~\cite{lynden2012baryonic}
\begin{align}
    M\lesssim\begin{cases}
        0.03M_\odot\left(\frac{a}{10^4\ \AU}\right)^{3}, &a\gtrsim 10^4\ \AU\\
        0.03M_\odot\left(\frac{a}{10^4\ \AU}\right)^{-1/2}, &5\times 10^3\ \AU\lesssim a\lesssim 10^{4}\ \AU\\
        0.04M_\odot\left(\frac{a}{5\times 10^3\ \AU}\right)^{-2}, &a\lesssim 5\times 10^3\ \AU~.
    \end{cases}
\end{align}
The distribution of the semimajor axes of observed Oort cloud comets additionally places the constraint~\cite{lynden2012baryonic}
\begin{align}
    M\lesssim 0.03M_\odot\left(\frac{a}{10^{3}\AU}\right),\quad a\lesssim 2\times 10^3\AU~.
\end{align}

\subsection{Solar System}

Dark masses in the Solar System can gravitationally alter the orbits of visible objects around the Sun, such as planets, asteroids, comets, and spacecraft. Based on the measured positions and velocities of these objects, future configurations of these bodies, called \textit{ephemerides}, could be predicted and compared with observations. Unmodeled dark masses would cause the ephemerides to fail at describing the observation data at some point. If the dark masses have a distribution that extends to the orbits of the Solar System objects, it would contribute non-inverse-square forces on these objects, which in turn cause compounding excess precession beyond known GR effects. The latter is observable as periastron drifts, whose non-observation leads to the constraints $\rho_{\rm dark}\lesssim 6\times 10^3, 8\times 10^3,8\times 10^4 \ \GeV/\cm^3$ at the orbital radii of Saturn, Mars, and the Earth. Translating these into limits on the total dark mass in the Solar System requires making assumption on the dark mass profile. Ref.~\cite{2013AstL...39..141P} considers an exponentially decaying profile $\rho_{\rm dark}\propto e^{-cr}$, while Ref.~\cite{Anderson:2020rdk} considers a power-law profile $\rho_{\rm dark}\propto r^{-n}$. If the dark mass is uniformly distributed, then the limit on the total dark mass in the Solar System is around $M\lesssim 10^{-13} \ M_\odot$.

We now estimate what would be the limit from ephemerides of inner planets if the dark mass is point-like. Since there seems to be no such analysis thus far, we provide our own crude estimate. Ephemerides limits are mainly driven by a few planets in the Solar System, in particular Earth, Mars, and Saturn. The limit on additional perihelion precession of Mars from Ref.~\cite{2013AstL...39..141P} is $d\theta/dt\lesssim 2\times 10^{-12}\ \text{rad}/\yr$. A dark mass $M$ with a velocity $\sim 10^{-4}$ (typical of Solar System orbits) passing by Mars with an impact parameter $d_{\rm min}$ would impart an extra angular displacement on Mars 
\begin{align}
    |\Delta\theta|\sim \frac{GM}{d_{\rm min}^2}\left(\frac{d_{\rm min}}{10^{-4}}\right)^2\times \frac{1}{a_{\rm Mars}}~,
\end{align}
where $a_{\rm Mars}=1.5\ \AU$ is the semimajor axis of Mars's orbit around the Sun. We assume such an encounter occurs at most once in a Mars revolution and require the unmodeled $|\Delta \theta|$ be less than $2\times 10^{-12}\ \text{rad}$, which amounts to
\begin{align}
    M\lesssim 3\times 10^{-12}M_\odot,\quad a\lesssim 2\ \AU~,
\end{align}
roughly independent of $d_{\rm min}$, as long as $d_{\rm min}\sim a\sim  \AU$. For larger values of $a$, ephemerides of outer planets sets the constraint~\cite{1991AJ....101.2274H}
\begin{align}
    M\lesssim 9\times 10^{-5} M_\odot\left(\frac{a}{100\AU}\right)^3,\quad a\gtrsim 2\ \AU~.
\end{align}
There are also constraints on the amount of dark mass residing outside the Earth but within the Moon's orbit around the Earth from combining lunar-, asteroid-, and satellite-ranging data~\cite{Adler:2008rq}
\begin{align}
    M\lesssim 1\times 10^{-14}M_\odot,\quad R_\oplus\lesssim \tilde{a}\lesssim 60R_\oplus ~,
\end{align}
where $\tilde{a}$ is the dark mass's semimajor axis from the center of the Earth.

Ephemerides of inner planets could also be used to measure the mass loss rate of the Sun. Ref.~\cite{2012SoSyR..46...78P} found $d\ln (G\dot{M}_\odot)/dt\approx -(5.0\pm 4.1)\times10^{-14}/\yr$, which is understood to be mainly due to the solar wind; see also Refs.~\cite{Pitjeva:2013xxa,Pitjeva:2021hnc}. If a total dark mass $M$ is present inside the orbits of the inner planets of the Solar System (i.e., $r\lesssim 2\ \AU$ from the Sun), the magnitude of its net mass loss rate must not exceed the measured rate, $|\dot{M}|/M_\odot\lesssim 10^{-13}/\yr$, assuming the time-variation of $G$ is not fine-tuned to cancel out that of $M$. Similarly, decades of lunar-ranging observations have measured the fractional time variation of the $GM$ of the Earth to be $d \, \text{ln} (G\dot{M}_\oplus)/dt=(2\pm 7)\times 10^{-13}/\yr$~\cite{Muller:2007zzb}. This translates to the rough constraint $|\dot{M}/M_\oplus|\lesssim 10^{-12}/\yr$ on the dark mass $M$ within the orbit of the Moon (i.e., $r\lesssim 60 \ R_\oplus$ from the Earth). Transient dark-mass losses could perhaps be constrained better than these time-averaged limits if they are specifically searched for in the data.

\subsection{Sun's and Earth's Interiors}
\label{app:interiors}

If the dark mass is located inside an astrophysical body in the Solar System, dark-mass limits from ephemerides become essentially non-existent due to degeneracy with the mass of the astrophysical body hosting it. While HFGW sources can hide in the Moon and other planets, we focus on the Sun and the Earth.

A point-like dark mass at rest at the center of the Sun is limited by helioseismology to be~\cite{Bahcall:2004qv,2013AstL...39..141P}
\begin{align}
    M\lesssim 1\%M_\odot,\quad\text{center of }\odot ~.
\end{align}
As we are not aware of prior work that places limits on \textit{static} dark masses residing inside the Earth, we conservatively require a dark mass at the center of the Earth to satisfy
\begin{align}
    M\lesssim 10\%M_\oplus,\quad \text{center of }\oplus~.
\end{align}
We expect a dedicated seismology analysis would give a far stronger limit. Besides seismology, gravitational measurements of the Earth's multipole moments by orbiting satellites would provide additional constraints~\cite{2024ESRv..25304783E}. Future atmospheric neutrino experiments may also probe the interior of the Earth, and hence the possible existence of dark masses~\cite{Rott:2015kwa,Winter:2015zwx,Denton:2021rgt,Kelly:2021jfs}.

A preexisting dark mass $M$ bound to the Sun/Earth may still be undergoing an orbital motion under the Sun's/Earth's gravitational potential today, with an orbital frequency $f\sim 0.3 \, \sqrt{G\rho_{\odot,\oplus}}$ which is about $0.9\ \text{mHz}$/$0.2\ \text{mHz}$ inside the Sun/Earth. The dark mass is more likely to stay dynamical at present if it arrived recently and has extremely tiny non-gravitational interactions with SM matter. Even if the dark mass interacts only gravitationally, it is subject to dynamical friction inside of the Earth/Sun via gravitational excitation of sound waves~\cite{Ostriker:1998fa,Genolini:2020ejw}
\begin{align}
    F_{\rm DF}\sim \frac{4\pi G^2M^2\rho}{v^2}\text{min}\left(10,\frac{v^3}{3c_s^3}\right) \label{eq:DF}~,
\end{align}
where we have assumed that the Coulomb logarithm yields $\ln\Lambda\sim 10$. This causes the orbit to decay over the timescale of 
\begin{align}
    \tau_{\rm DF}\sim \frac{Mv}{F_{\rm DF}}\sim\begin{cases}
     1\ \text{Gyr}\left(\frac{M}{10^{-16}M_\odot}\right)^{-1}, &\odot\ \text{interior}\\
     1\ \text{Gyr}\left(\frac{M}{10^{-19}M_\odot}\right)^{-1}, &\oplus\ \text{interior}~,
    \end{cases}
    \label{eq:DFtime}
\end{align}
for $v\ll c_{s,\oplus}\sim 3\times 10^{-5}, c_{s,\odot}\sim 1\times 10^{-3}$, $\rho_\oplus\sim 10\ \text{g}/\cm^3$, and $\rho_{\odot}\sim 150\ \text{g}/\cm^3$. If dynamical friction fails to damp the bulk motion of the dark mass at present, it can be probed with a network of superconducting gravimeters placed at various locations on Earth. Decade's worth of data taken by these gravimeters have already placed the following limits on a dark mass orbiting the Earth~\cite{Namigata:2022vry,Horowitz:2019fus}
\begin{align}
    M_{\oplus\text{-bound}}\lesssim 1\times 10^{-18}M_\odot\left(\frac{\tilde{a}}{R_\oplus}\right)^5,\quad 0.1R_\oplus\lesssim \tilde{a}\lesssim7R_\oplus~.
\end{align}
Here, the dark mass is assumed to be moving in an elliptical orbit with a semimajor axis $\tilde{a}$ from the center of the Earth. This includes orbits that are entirely inside the Earth as well as those that emerge outside of the Earth's surface. The above expression, however, does not capture how the gravimeters loses sensitivity for $\tilde{a}\lesssim 0.1 \ R_\oplus$.

\section{Gravitational Memory}

\label{app:memory}

The time-dependent memory strain $h_{ij}^{\rm mem}(t)$ at a distance $d$ from a source emitting GWs with luminosity $\mathcal{L}_{\rm GW}$ is given by Thorne's formula~\cite{Thorne:1992sdb}
    \begin{align}
        \left[h_{ij}^{\rm mem}\right]_{\rm TT}=\frac{4G}{d}\int_{-\infty}^tdt'\int d\Omega' \frac{d \mathcal{L}_\text{GW}}{d\Omega'}\left[\frac{\hat{n}'_i\hat{n}'_j}{1-\hat{n}\cdot\hat{n}'}\right]_{\rm TT}~,
    \end{align}  
where $\hat{n}$ the unit vector from the source to a point whose strain if of interest, $\hat{n}'$ is a unit vector toward a differential patch of the sky with solid angle $d\Omega'$, and the subscript $_{\rm TT}$ means ``keep only the transverse and traceless part." This integral is sensitive to the entire history of the source's GW emission. The order of magnitude of the memory strain signal is $h_{ij}^{\rm mem}\sim GE_\text{GW}/d$, with $E_{\rm GW}\sim \int _{-\infty}^td t'\,\mathcal{L}_{\rm GW}(t')$ the total energy emitted by the source up to time $t$.

For a HFGW source with a GW-emitting mass $M$ and size $L$, producing a Dirac-delta-like GW burst with a typical GW frequency $f_{\rm HF}$, the $h_{ij}^{\rm mem}(t)$ looks like a step function that turns on at the moment of arrival of the emission, $h_{\rm mem}(t)\sim (GE_\text{GW}/d)\Theta(t-t_\text{GW})$, where $E_\text{GW}\sim ML^2f_\text{HF}^2$ if all the internal kinetic energy of the source is emitted in the GW burst. If the GW burst has a finite duration $t_\text{sig}$, the memory step would have a finite turn-on time matching the timescale of the burst, and consequently, the Fourier transform of the memory signal $\tilde{h}_{ij}^{\rm mem}(f)$ has support only for $f\lesssim f_\text{max}^{\rm mem}\sim t_\text{sig}^{-1}$. In the low-frequency limit, $f\ll t_\text{sig}^{-1}$, the $\tilde{h}_{ij}^{\rm mem}(f)$ has a universal $f^{-1}$ scaling
\begin{align}
    \tilde{h}_{ij}^{\rm mem}(f)\sim \frac{GML^2f_\text{HF}^2}{d}\frac{1}{2\pi f}~.
\end{align}
Memory signals have been searched for in LIGO~\cite{Ebersold:2020zah} and NANOGrav~\cite{Agazie:2025oug,NANOGrav:2019vto}, finding null results.   Setting $f=10\ \Hz$ and $S_n^\text{LIGO}(10\ \Hz)\sim 10^{-46}\ \Hz^{-1}$, we find that the memory signal would be seen at LIGO/Virgo with
\begin{align}
   \text{SNR}_{\rm mem}\sim 2\times 10^{8}\left(\frac{M}{M_\odot}\right)\left(\frac{L}{\text{m}}\right)^2\left(\frac{f_\text{HF}}{\MHz}\right)^2\left(\frac{d}{\ \AU}\right)^{-1}~.
\end{align}
For a PBH merger, the parameters are tied as $L\sim GM\sim t_\text{sig}\sim f_\text{HF}^{-1}$, and so $\text{SNR}_{\rm mem}\sim \,(f_\text{HF}/\MHz)^{-1}(d/10\ \text{Mpc})^{-1}$. The same merger would produce a signal in a $S_n=10^{-40}\ \Hz^{-1}$ HFGW detector with $\text{SNR}_\text{HF}\sim 10^{-8}(f/\MHz)^{-3/2}(d/10\ \text{Mpc})^{-1}$. Thus, the ratio $\text{SNR}_\text{HF}/\text{SNR}_{\rm mem}\sim 10^{-8} (f_\text{HF}/10\ \kHz)^{-1/2}$ is always less than unity for $f_\text{HF}\gtrsim 10\ \kHz$. In other words, a hypothetical PBH merger that would leave a detectable signal in the assumed HFGW detector is already ruled out by the non-observation of memory signals in the LIGO/Virgo band~\cite{McNeill:2017uvq}.

Having said that, limits from LIGO/Virgo can be avoided if $f_\text{max}^{\rm mem}\sim t_\text{sig}^{-1}\lesssim f_{\rm LIGO,min}\approx 10\ \Hz$, i.e., if $t_\text{sig}\gtrsim 0.1\ \text{s}$, in which case the memory signal in LIGO/Virgo would be buried in seismic noise. Pulsar-timing arrays (PTAs) are in principle sensitive to such low-frequency memory signals, however the time-delay effects due to local sources in the Solar System is heavily suppressed by pulsar distances. Thus, the resulting limits set by PTAs are very weak. Local and shorter baseline experiments may set stronger constraints. Lunar-ranging measurements (e.g., of the APOLLO station) and satellite-ranging measurements (e.g., of the LAGEOS satellite) should, in principle, have sensitivity to memory signals. However, there has not been a dedicated analysis to search for memory signals in the existing data from these instruments.

An ellipsoidal spinning football emits GWs with a luminosity given by
\begin{align}
    \mathcal{L}_\text{GW}\sim GM^2L^4f(t)^6~,
\end{align}
where $L^2=L_1^2-L_2^2$ and $L_{1,2}$ are the two semiaxes of the football. In response to the GW emission, the football's frequency evolves as
$f(t)=f_0(1+t/\tau_0)^{-1/4}$, where $f_0$ is the initial GW frequency sourced by the football and $\tau_0\sim (GML^2f_0^4)^{-1}$. Assuming the football arrives at a distance $d$ or starts rotating at $t=0$, the memory signal is given by the integral $h_{\rm mem}(t)\sim (4G/d)\int_{0}^tdt' \mathcal{L}_\text{GW}(t') $, which yields
\begin{align}
    h_{\rm mem}(t)\sim \frac{GML^2f_0^2}{d} \left[1-\left(1+\frac{t}{\tau_0}\right)^{-1/2}\right]~.
\end{align}
Note that in terms of strength this memory strain is comparable to that sourced by a merging blob with similar $M,L,f,d$. However, unlike a merging blob which yields a step-function-like $h_{\rm mem}(t)$, a down-spinning football has a $h_{\rm mem}(t)$ that is gradually increasing in strength over time. Its Fourier transform over the observation time window of an experiment, $[0,t_\text{obs}]$, for $f\gtrsim t_\text{obs}^{-1},\tau_0^{-1}$, is oscillatory in $f$ with an envelope given by
\begin{align}
    |\tilde{h}_{\rm mem}(f)|\sim \frac{GML^2f_0^2}{d}\begin{cases}
        \frac{1}{2\pi f}, &\tau_0\ll t_\text{obs}\\
        \frac{1}{2\pi f}\frac{t_{\rm obs}}{\tau_0}, &\tau_0\gg t_\text{obs}
    \end{cases}~.
\end{align}
If we conservatively assume that the initial frequency $f_0$ directly falls in an HFGW band of interest, $f_0\sim f_\text{HF}$, the resulting memory signal is comparable to that of a merging binary with similar $M,L,f_\text{HF}$ for $\tau_0\sim t_\text{obs}$, but could be suppressed for $\tau_0\gg t_\text{obs}$. The similarity between the $h_{\rm mem}(t)$ of a spinning football with that of a merging binary may be understood by interpreting the envelope of the football's luminosity $\mathcal{L}_\text{GW}$ as the growing part of a burst and the time cutoff set by the finite observation time as the decaying part of a burst.

%In principle, there should also be the rising part of the $\mathcal{L}_\text{GW}$ which is to do with the football's arrival history, which we did not consider.

\bibliography{references}

@article{Christodoulou:1991cr,
    author = "Christodoulou, D.",
    title = "{Nonlinear nature of gravitation and gravitational wave experiments}",
    doi = "10.1103/PhysRevLett.67.1486",
    journal = "Phys. Rev. Lett.",
    volume = "67",
    pages = "1486--1489",
    year = "1991"
}

@book{Raffelt:1996wa,
    author = "Raffelt, G. G.",
    title = "{Stars as laboratories for fundamental physics}: {The astrophysics of neutrinos, axions, and other weakly interacting particles}",
    isbn = "978-0-226-70272-8",
    month = "5",
    year = "1996"
}

@article{Safdi:2022xkm,
    author = "Safdi, Benjamin R.",
    title = "{TASI Lectures on the Particle Physics and Astrophysics of Dark Matter}",
    eprint = "2303.02169",
    archivePrefix = "arXiv",
    primaryClass = "hep-ph",
    doi = "10.22323/1.439.0009",
    journal = "PoS",
    volume = "TASI2022",
    pages = "009",
    year = "2024"
}

@article{Coccia:2004gw,
    author = "Coccia, Eugenio and Dubath, Florian and Maggiore, Michele",
    title = "{On the possible sources of gravitational wave bursts detectable today}",
    eprint = "gr-qc/0405047",
    archivePrefix = "arXiv",
    doi = "10.1103/PhysRevD.70.084010",
    journal = "Phys. Rev. D",
    volume = "70",
    pages = "084010",
    year = "2004"
}

@article{LIGOScientific:2016aoc,
    author = "Abbott, B. P. and others",
    collaboration = "LIGO Scientific, Virgo",
    title = "{Observation of Gravitational Waves from a Binary Black Hole Merger}",
    eprint = "1602.03837",
    archivePrefix = "arXiv",
    primaryClass = "gr-qc",
    reportNumber = "LIGO-P150914",
    doi = "10.1103/PhysRevLett.116.061102",
    journal = "Phys. Rev. Lett.",
    volume = "116",
    number = "6",
    pages = "061102",
    year = "2016"
}

@article{Muller:2007zzb,
    author = "Muller, Jurgen and Biskupek, Liliane",
    title = "{Variations of the gravitational constant from lunar laser ranging data}",
    doi = "10.1088/0264-9381/24/17/017",
    journal = "Class. Quant. Grav.",
    volume = "24",
    pages = "4533--4538",
    year = "2007"
}

@article{Pitjeva:2013xxa,
    author = "Pitjeva, E. V. and Pitjev, N. P.",
    title = "{Relativistic effects and dark matter in the Solar system from observations of planets and spacecraft}",
    eprint = "1306.3043",
    archivePrefix = "arXiv",
    primaryClass = "astro-ph.EP",
    doi = "10.1093/mnras/stt695",
    journal = "Mon. Not. Roy. Astron. Soc.",
    volume = "432",
    pages = "3431",
    year = "2013"
}

@article{Holometer:2016qoh,
    author = "Chou, Aaron S. and others",
    collaboration = "Holometer",
    title = "{MHz Gravitational Wave Constraints with Decameter Michelson Interferometers}",
    eprint = "1611.05560",
    archivePrefix = "arXiv",
    primaryClass = "astro-ph.IM",
    reportNumber = "FERMILAB-PUB-16-449-AE",
    doi = "10.1103/PhysRevD.95.063002",
    journal = "Phys. Rev. D",
    volume = "95",
    number = "6",
    pages = "063002",
    year = "2017"
}

@book{lynden2012baryonic,
  title={Baryonic dark matter},
  author={Lynden-Bell, Donald and Gilmore, Gerry},
  volume={306},
  year={2012},
  publisher={Springer Science \& Business Media}
}

@ARTICLE{1991AJ....101.2274H,
       author = {{Hogg}, David W. and {Quinlan}, Gerald D. and {Tremaine}, Scott},
        title = "{Dynamical Limits on Dark Mass in the Solar System}",
      journal = {\aj},
         year = 1991,
        month = jun,
       volume = {101},
        pages = {2274},
          doi = {10.1086/115849},
       adsurl = {https://ui.adsabs.harvard.edu/abs/1991AJ....101.2274H}
}

@article{Aggarwal:2020olq,
    author = "Aggarwal, Nancy and others",
    title = "{Challenges and opportunities of gravitational-wave searches at MHz to GHz frequencies}",
    eprint = "2011.12414",
    archivePrefix = "arXiv",
    primaryClass = "gr-qc",
    reportNumber = "CERN-TH-2020-185, HIP-2020-28/TH, DESY 20-195, CERN-TH-2020-185, HIP-2020-28/TH, DESY 20-195",
    doi = "10.1007/s41114-021-00032-5",
    journal = "Living Rev. Rel.",
    volume = "24",
    number = "1",
    pages = "4",
    year = "2021"
}

@article{Aggarwal:2025noe,
    author = "Aggarwal, Nancy and others",
    title = "{Challenges and Opportunities of Gravitational Wave Searches above 10 kHz}",
    eprint = "2501.11723",
    archivePrefix = "arXiv",
    primaryClass = "gr-qc",
    reportNumber = "CERN-TH-2025-014, DESY-25-007",
    month = "1",
    year = "2025"
}

@article{Bai:2024pki,
    author = "Bai, Yang and Lu, Sida and Orlofsky, Nicholas",
    title = "{Gravitational waves from dark binaries with finite-range dark forces}",
    eprint = "2412.15158",
    archivePrefix = "arXiv",
    primaryClass = "gr-qc",
    doi = "10.1088/1475-7516/2025/03/010",
    journal = "JCAP",
    volume = "03",
    pages = "010",
    year = "2025"
}

@article{Kaplan:2024dsn,
    author = "Kaplan, David E. and Luo, Xuheng and Nguyen, Ngan H. and Rajendran, Surjeet and Tanin, Erwin H.",
    title = "{Indirect searches for ultraheavy dark matter in the time domain}",
    eprint = "2407.06262",
    archivePrefix = "arXiv",
    primaryClass = "hep-ph",
    reportNumber = "FERMILAB-PUB-24-0484-SQMS-V",
    doi = "10.1103/PhysRevD.111.023041",
    journal = "Phys. Rev. D",
    volume = "111",
    number = "2",
    pages = "023041",
    year = "2025"
}

@article{Bai:2023lyf,
    author = "Bai, Yang and Lu, Sida and Orlofsky, Nicholas",
    title = "{Gravitational waves from more attractive dark binaries}",
    eprint = "2312.13378",
    archivePrefix = "arXiv",
    primaryClass = "astro-ph.CO",
    doi = "10.1088/1475-7516/2024/08/057",
    journal = "JCAP",
    volume = "08",
    pages = "057",
    year = "2024"
}

@article{Fedderke:2024hfy,
    author = "Fedderke, Michael A. and Kaplan, David E. and Mathur, Anubhav and Rajendran, Surjeet and Tanin, Erwin H.",
    title = "{Fireball antinucleosynthesis}",
    eprint = "2402.15581",
    archivePrefix = "arXiv",
    primaryClass = "hep-ph",
    reportNumber = "FERMILAB-PUB-24-0514-SQMS-V",
    doi = "10.1103/PhysRevD.109.123028",
    journal = "Phys. Rev. D",
    volume = "109",
    number = "12",
    pages = "123028",
    year = "2024"
}

@article{Figueroa:2019paj,
    author = "Figueroa, Daniel G. and Tanin, Erwin H.",
    title = "{Ability of LIGO and LISA to probe the equation of state of the early Universe}",
    eprint = "1905.11960",
    archivePrefix = "arXiv",
    primaryClass = "astro-ph.CO",
    doi = "10.1088/1475-7516/2019/08/011",
    journal = "JCAP",
    volume = "08",
    pages = "011",
    year = "2019"
}

@article{Gatti:2024mde,
    author = "Gatti, Claudio and Visinelli, Luca and Zantedeschi, Michael",
    title = "{Cavity detection of gravitational waves: Where do we stand?}",
    eprint = "2403.18610",
    archivePrefix = "arXiv",
    primaryClass = "gr-qc",
    reportNumber = "CA21106; CA21136",
    doi = "10.1103/PhysRevD.110.023018",
    journal = "Phys. Rev. D",
    volume = "110",
    number = "2",
    pages = "023018",
    year = "2024"
}

@article{Cruise:2012zz,
    author = "Cruise, A. M.",
    title = "{The potential for very high-frequency gravitational wave detection}",
    doi = "10.1088/0264-9381/29/9/095003",
    journal = "Class. Quant. Grav.",
    volume = "29",
    pages = "095003",
    year = "2012"
}

@article{Favata:2010zu,
    author = "Favata, Marc",
    editor = "Marka, Zsuzsa and Marka, Szabolcs",
    title = "{The gravitational-wave memory effect}",
    eprint = "1003.3486",
    archivePrefix = "arXiv",
    primaryClass = "gr-qc",
    doi = "10.1088/0264-9381/27/8/084036",
    journal = "Class. Quant. Grav.",
    volume = "27",
    pages = "084036",
    year = "2010"
}

@article{McNeill:2017uvq,
    author = "McNeill, Lucy O. and Thrane, Eric and Lasky, Paul D.",
    title = "{Detecting Gravitational Wave Memory without Parent Signals}",
    eprint = "1702.01759",
    archivePrefix = "arXiv",
    primaryClass = "astro-ph.IM",
    doi = "10.1103/PhysRevLett.118.181103",
    journal = "Phys. Rev. Lett.",
    volume = "118",
    number = "18",
    pages = "181103",
    year = "2017"
}

@article{Thorne:1992sdb,
    author = "Thorne, Kip S.",
    title = "{Gravitational-wave bursts with memory: The Christodoulou effect}",
    doi = "10.1103/PhysRevD.45.520",
    journal = "Phys. Rev. D",
    volume = "45",
    number = "2",
    pages = "520--524",
    year = "1992"
}

@article{Planck:2018vyg,
    author = "Aghanim, N. and others",
    collaboration = "Planck",
    title = "{Planck 2018 results. VI. Cosmological parameters}",
    eprint = "1807.06209",
    archivePrefix = "arXiv",
    primaryClass = "astro-ph.CO",
    doi = "10.1051/0004-6361/201833910",
    journal = "Astron. Astrophys.",
    volume = "641",
    pages = "A6",
    year = "2020",
    note = "[Erratum: Astron.Astrophys. 652, C4 (2021)]"
}

@article{Genolini:2020ejw,
    author = "G{\'e}nolini, Yoann and Serpico, Pasquale and Tinyakov, Peter",
    title = "{Revisiting primordial black hole capture into neutron stars}",
    eprint = "2006.16975",
    archivePrefix = "arXiv",
    primaryClass = "astro-ph.HE",
    doi = "10.1103/PhysRevD.102.083004",
    journal = "Phys. Rev. D",
    volume = "102",
    number = "8",
    pages = "083004",
    year = "2020"
}

@article{Bahcall:2000nu,
    author = "Bahcall, John N. and Pinsonneault, M. H. and Basu, Sarbani",
    title = "{Solar models: Current epoch and time dependences, neutrinos, and helioseismological properties}",
    eprint = "astro-ph/0010346",
    archivePrefix = "arXiv",
    doi = "10.1086/321493",
    journal = "Astrophys. J.",
    volume = "555",
    pages = "990--1012",
    year = "2001"
}

@ARTICLE{2016SciA....2E0802S,
       author = {{Sakamaki}, Tatsuya and {Ohtani}, Eiji and {Fukui}, Hiroshi and {Kamada}, Seiji and {Takahashi}, Suguru and {Sakairi}, Takanori and {Takahata}, Akihiro and {Sakai}, Takeshi and {Tsutsui}, Satoshi and {Ishikawa}, Daisuke and {Shiraishi}, Rei and {Seto}, Yusuke and {Tsuchiya}, Taku and {Baron}, Alfred Q.~R.},
        title = "{Constraints on Earth's inner core composition inferred from measurements of the sound velocity of hcp-iron in extreme conditions}",
      journal = {Science Advances},
         year = 2016,
        month = feb,
       volume = {2},
       number = {2},
        pages = {e1500802-e1500802},
          doi = {10.1126/sciadv.1500802},
       adsurl = {https://ui.adsabs.harvard.edu/abs/2016SciA....2E0802S}
}

@ARTICLE{2013AstL...39..141P,
       author = {{Pitjev}, N.~P. and {Pitjeva}, E.~V.},
        title = "{Constraints on dark matter in the solar system}",
      journal = {Astronomy Letters},
         year = 2013,
        month = mar,
       volume = {39},
       number = {3},
        pages = {141-149},
          doi = {10.1134/S1063773713020060},
archivePrefix = {arXiv},
       eprint = {1306.5534},
 primaryClass = {astro-ph.EP},
       adsurl = {https://ui.adsabs.harvard.edu/abs/2013AstL...39..141P}
}

@article{Anderson:2020rdk,
    author = "Anderson, Noah B. and Partenheimer, Angelina and Wiser, Timothy D.",
    title = "{Direct detection signatures of a primordial Solar dark matter halo}",
    eprint = "2007.11016",
    archivePrefix = "arXiv",
    primaryClass = "hep-ph",
    month = "7",
    year = "2020"
}

@article{Adler:2008rq,
    author = "Adler, Stephen L.",
    title = "{Placing direct limits on the mass of earth-bound dark matter}",
    eprint = "0808.0899",
    archivePrefix = "arXiv",
    primaryClass = "astro-ph",
    doi = "10.1088/1751-8113/41/41/412002",
    journal = "J. Phys. A",
    volume = "41",
    pages = "412002",
    year = "2008"
}

@article{Bahcall:2004qv,
    author = "Bahcall, John N.",
    editor = {Bergstr{\"o}m, L. and Botner, O. and Carlson, P. and Hulth, P. O. and Ohlsson, T.},
    title = "{Solar models and solar neutrinos: Current status}",
    eprint = "hep-ph/0412068",
    archivePrefix = "arXiv",
    doi = "10.1088/0031-8949/2005/T121/006",
    journal = "Phys. Scripta T",
    volume = "121",
    pages = "46--50",
    year = "2005"
}

@article{Rott:2015kwa,
    author = "Rott, Carsten and Taketa, Akimichi and Bose, Debanjan",
    title = "{Spectrometry of the Earth using Neutrino Oscillations}",
    eprint = "1502.04930",
    archivePrefix = "arXiv",
    primaryClass = "physics.geo-ph",
    doi = "10.1038/srep15225",
    journal = "Sci. Rep.",
    volume = "5",
    pages = "15225",
    year = "2015"
}

@article{Winter:2015zwx,
    author = "Winter, Walter",
    title = "{Atmospheric neutrino oscillations for Earth tomography}",
    eprint = "1511.05154",
    archivePrefix = "arXiv",
    primaryClass = "hep-ph",
    reportNumber = "DESY-15-227",
    doi = "10.1016/j.nuclphysb.2016.03.033",
    journal = "Nucl. Phys. B",
    volume = "908",
    pages = "250--267",
    year = "2016"
}

@article{Denton:2021rgt,
    author = "Denton, Peter B. and Pestes, Rebekah",
    title = "{Neutrino oscillations through the Earth{\textquoteright}s core}",
    eprint = "2110.01148",
    archivePrefix = "arXiv",
    primaryClass = "hep-ph",
    doi = "10.1103/PhysRevD.104.113007",
    journal = "Phys. Rev. D",
    volume = "104",
    number = "11",
    pages = "113007",
    year = "2021"
}

@article{Kelly:2021jfs,
    author = "Kelly, Kevin J. and Machado, Pedro A. N. and Martinez-Soler, Ivan and Perez-Gonzalez, Yuber F.",
    title = "{DUNE atmospheric neutrinos: Earth tomography}",
    eprint = "2110.00003",
    archivePrefix = "arXiv",
    primaryClass = "hep-ph",
    reportNumber = "FERMILAB-PUB-21-459-T, NUHEP-TH/21-15",
    doi = "10.1007/JHEP05(2022)187",
    journal = "JHEP",
    volume = "05",
    pages = "187",
    year = "2022"
}

@article{Ostriker:1998fa,
    author = "Ostriker, Eve C.",
    title = "{Dynamical friction in a gaseous medium}",
    eprint = "astro-ph/9810324",
    archivePrefix = "arXiv",
    doi = "10.1086/306858",
    journal = "Astrophys. J.",
    volume = "513",
    pages = "252",
    year = "1999"
}

@article{Namigata:2022vry,
    author = "Namigata, Tomoyo and Horowitz, C. J. and Widmer-Schnidrig, R.",
    title = "{Gravitational search for near Earth black holes or other compact dark objects}",
    eprint = "2201.06511",
    archivePrefix = "arXiv",
    primaryClass = "gr-qc",
    month = "1",
    year = "2022"
}

@article{Horowitz:2019fus,
    author = "Horowitz, C. J. and Widmer-Schnidrig, R.",
    title = "{Gravimeter search for compact dark matter objects moving in the Earth}",
    eprint = "1912.00940",
    archivePrefix = "arXiv",
    primaryClass = "astro-ph.EP",
    doi = "10.1103/PhysRevLett.124.051102",
    journal = "Phys. Rev. Lett.",
    volume = "124",
    number = "5",
    pages = "051102",
    year = "2020"
}

@ARTICLE{2024ESRv..25304783E,
       author = {{Eshagh}, Mehdi and {Jin}, Shuanggen and {Pail}, Roland and {Barzaghi}, Riccardo and {Tsoulis}, Dimitrios and {Tenzer}, Robert and {Nov{\'a}k}, Pavel},
        title = "{Satellite gravimetry: Methods, products, applications, and future trends}",
      journal = {Earth Science Reviews},
         year = 2024,
        month = jun,
       volume = {253},
          eid = {104783},
        pages = {104783},
          doi = {10.1016/j.earscirev.2024.104783},
       adsurl = {https://ui.adsabs.harvard.edu/abs/2024ESRv..25304783E}
}

@article{Ebersold:2020zah,
    author = "Ebersold, Michael and Tiwari, Shubhanshu",
    title = "{Search for nonlinear memory from subsolar mass compact binary mergers}",
    eprint = "2005.03306",
    archivePrefix = "arXiv",
    primaryClass = "gr-qc",
    doi = "10.1103/PhysRevD.101.104041",
    journal = "Phys. Rev. D",
    volume = "101",
    number = "10",
    pages = "104041",
    year = "2020"
}

@article{Agazie:2025oug,
    author = "Agazie, Gabriella and others",
    title = "{The NANOGrav 15 yr Data Set: Search for Gravitational-wave Memory}",
    eprint = "2502.18599",
    archivePrefix = "arXiv",
    primaryClass = "gr-qc",
    doi = "10.3847/1538-4357/add874",
    journal = "Astrophys. J.",
    volume = "987",
    number = "1",
    pages = "5",
    year = "2025"
}

@article{NANOGrav:2019vto,
    author = "Aggarwal, K. and others",
    collaboration = "NANOGrav",
    title = "{The NANOGrav 11 yr Data Set: Limits on Gravitational Wave Memory}",
    eprint = "1911.08488",
    archivePrefix = "arXiv",
    primaryClass = "astro-ph.HE",
    doi = "10.3847/1538-4357/ab6083",
    journal = "Astrophys. J.",
    volume = "889",
    pages = "38",
    year = "2020"
}

@ARTICLE{2012SoSyR..46...78P,
       author = {{Pitjeva}, E.~V. and {Pitjev}, N.~P.},
        title = "{Changes in the Sun's mass and gravitational constant estimated using modern observations of planets and spacecraft}",
      journal = {Solar System Research},
         year = 2012,
        month = feb,
       volume = {46},
       number = {1},
        pages = {78-87},
          doi = {10.1134/S0038094612010054},
archivePrefix = {arXiv},
       eprint = {1108.0246},
 primaryClass = {astro-ph.SR},
       adsurl = {https://ui.adsabs.harvard.edu/abs/2012SoSyR..46...78P}
}

@article{Pitjeva:2021hnc,
    author = "Pitjeva, E. V. and Pitjev, N. P. and Pavlov, D. A. and Turygin, C. C.",
    title = "{Estimates of the change rate of solar mass and gravitational constant based on the dynamics of the Solar System}",
    eprint = "2201.09804",
    archivePrefix = "arXiv",
    primaryClass = "astro-ph.EP",
    doi = "10.1051/0004-6361/202039893",
    journal = "Astron. Astrophys.",
    volume = "647",
    pages = "A141",
    year = "2021"
}

@ARTICLE{1985Ap&SS.111..335B,
       author = {{Beltrami}, H. and {Chau}, W.~Y.},
        title = "{Analytic Result for the Properties of Gravitational Waves Emitted by a Large Class of Model Sources}",
      journal = {\apss},
         year = 1985,
        month = apr,
       volume = {111},
       number = {2},
        pages = {335-341},
          doi = {10.1007/BF00649973},
       adsurl = {https://ui.adsabs.harvard.edu/abs/1985Ap&SS.111..335B}
}

@article{Wang:2025mea,
    author = "Wang, Xinpeng and Lu, Yifan and Picker, Zachary S. C. and Kusenko, Alexander and Sasaki, Misao",
    title = "{When Tiny Halos Stir Spacetime: Gravitational Waves from Fifth-Force Mergers}",
    eprint = "2510.12984",
    archivePrefix = "arXiv",
    primaryClass = "astro-ph.CO",
    month = "10",
    year = "2025"
}

@article{Fan:2013yva,
    author = "Fan, JiJi and Katz, Andrey and Randall, Lisa and Reece, Matthew",
    title = "{Double-Disk Dark Matter}",
    eprint = "1303.1521",
    archivePrefix = "arXiv",
    primaryClass = "astro-ph.CO",
    doi = "10.1016/j.dark.2013.07.001",
    journal = "Phys. Dark Univ.",
    volume = "2",
    pages = "139--156",
    year = "2013"
}

@article{Ebadi:2025umm,
    author = "Ebadi, Reza and Tanin, Erwin H.",
    title = "{Making the Subdominant Dominant: Gravothermal Pile-Up of Collisional Dark Matter Around Compact Objects}",
    eprint = "2507.11601",
    archivePrefix = "arXiv",
    primaryClass = "hep-ph",
    month = "7",
    year = "2025"
}

@article{Ebadi:2021cte,
    author = "Ebadi, Reza and others",
    title = "{Ultraheavy dark matter search with electron microscopy of geological quartz}",
    eprint = "2105.03998",
    archivePrefix = "arXiv",
    primaryClass = "hep-ph",
    doi = "10.1103/PhysRevD.104.015041",
    journal = "Phys. Rev. D",
    volume = "104",
    number = "1",
    pages = "015041",
    year = "2021"
}

@article{Jacobs:2014yca,
    author = "Jacobs, David M. and Starkman, Glenn D. and Lynn, Bryan W.",
    title = "{Macro Dark Matter}",
    eprint = "1410.2236",
    archivePrefix = "arXiv",
    primaryClass = "astro-ph.CO",
    doi = "10.1093/mnras/stv774",
    journal = "Mon. Not. Roy. Astron. Soc.",
    volume = "450",
    number = "4",
    pages = "3418--3430",
    year = "2015"
}

@article{Kurita:2015vga,
    author = "Kurita, Yasunari and Nakano, Hiroyuki",
    title = "{Gravitational waves from dark matter collapse in a star}",
    eprint = "1510.00893",
    archivePrefix = "arXiv",
    primaryClass = "gr-qc",
    doi = "10.1103/PhysRevD.93.023508",
    journal = "Phys. Rev. D",
    volume = "93",
    number = "2",
    pages = "023508",
    year = "2016"
}

@ARTICLE{1971ApJ...166..175I,
       author = {{Ipser}, James R.},
        title = "{Gravitational Radiation from Slowly Rotating, Fully Relativistic Stars}",
      journal = {\apj},
         year = 1971,
        month = may,
       volume = {166},
        pages = {175},
          doi = {10.1086/150948},
       adsurl = {https://ui.adsabs.harvard.edu/abs/1971ApJ...166..175I}
}

@ARTICLE{1985ApJ...296..679P,
       author = {{Press}, W.~H. and {Spergel}, D.~N.},
        title = "{Capture by the sun of a galactic population of weakly interacting, massive particles}",
      journal = {\apj},
         year = 1985,
        month = sep,
       volume = {296},
        pages = {679-684},
          doi = {10.1086/163485},
       adsurl = {https://ui.adsabs.harvard.edu/abs/1985ApJ...296..679P}
}

@article{Goldman:1989nd,
    author = "Goldman, I. and Nussinov, S.",
    title = "{Weakly Interacting Massive Particles and Neutron Stars}",
    doi = "10.1103/PhysRevD.40.3221",
    journal = "Phys. Rev. D",
    volume = "40",
    pages = "3221--3230",
    year = "1989"
}

@ARTICLE{1990PhLB..238..337G,
       author = {{Gould}, Andrew and {Draine}, Bruce T. and {Romani}, Roger W. and {Nussinov}, Shmuel},
        title = "{Neutron stars: Graveyard of charged dark matter}",
      journal = {Physics Letters B},
         year = 1990,
        month = apr,
       volume = {238},
       number = {2-4},
        pages = {337-343},
          doi = {10.1016/0370-2693(90)91745-W},
       adsurl = {https://ui.adsabs.harvard.edu/abs/1990PhLB..238..337G}
}

@ARTICLE{2002ApJ...565..430F,
       author = {{Fryer}, Chris L. and {Holz}, Daniel E. and {Hughes}, Scott A.},
        title = "{Gravitational Wave Emission from Core Collapse of Massive Stars}",
      journal = {\apj},
         year = 2002,
        month = jan,
       volume = {565},
       number = {1},
        pages = {430-446},
          doi = {10.1086/324034},
archivePrefix = {arXiv},
       eprint = {astro-ph/0106113},
 primaryClass = {astro-ph},
       adsurl = {https://ui.adsabs.harvard.edu/abs/2002ApJ...565..430F}
}

@article{TitoDAgnolo:2024res,
    author = "Tito D'Agnolo, Raffaele and Ellis, Sebastian A. R.",
    title = "{Classical (and quantum) heuristics for gravitational wave detection}",
    eprint = "2412.17897",
    archivePrefix = "arXiv",
    primaryClass = "gr-qc",
    doi = "10.1007/JHEP04(2025)164",
    journal = "JHEP",
    volume = "04",
    pages = "164",
    year = "2025"
}

@article{Moore:2014lga,
    author = "Moore, C. J. and Cole, R. H. and Berry, C. P. L.",
    title = "{Gravitational-wave sensitivity curves}",
    eprint = "1408.0740",
    archivePrefix = "arXiv",
    primaryClass = "gr-qc",
    reportNumber = "LIGO-P1400129",
    doi = "10.1088/0264-9381/32/1/015014",
    journal = "Class. Quant. Grav.",
    volume = "32",
    number = "1",
    pages = "015014",
    year = "2015"
}

@article{Domcke:2024mfu,
    author = "Domcke, Valerie and Ellis, Sebastian A. R. and Rodd, Nicholas L.",
    title = "{Magnets are Weber Bar Gravitational Wave Detectors}",
    eprint = "2408.01483",
    archivePrefix = "arXiv",
    primaryClass = "hep-ph",
    reportNumber = "CERN-TH-2024-132",
    doi = "10.1103/966v-r5fm",
    journal = "Phys. Rev. Lett.",
    volume = "134",
    number = "23",
    pages = "231401",
    year = "2025"
}

@article{Schnabel:2024hem,
    author = "Schnabel, Roman and Korobko, Mikhail",
    title = "{Optical sensitivities of current gravitational wave observatories at higher kHz, MHz and GHz frequencies}",
    eprint = "2409.03019",
    archivePrefix = "arXiv",
    primaryClass = "astro-ph.IM",
    doi = "10.1038/s41598-025-08668-x",
    journal = "Sci. Rep.",
    volume = "15",
    number = "1",
    pages = "25733",
    year = "2025"
}

@article{Domcke:2024eti,
    author = "Domcke, Valerie and Ellis, Sebastian A. R. and Kopp, Joachim",
    title = "{Dielectric haloscopes as gravitational wave detectors}",
    eprint = "2409.06462",
    archivePrefix = "arXiv",
    primaryClass = "hep-ph",
    reportNumber = "CERN-TH-2024-151",
    doi = "10.1103/PhysRevD.111.035031",
    journal = "Phys. Rev. D",
    volume = "111",
    number = "3",
    pages = "035031",
    year = "2025"
}

@article{DeMiguel:2023nmz,
    author = "De Miguel, Javier and Hern{\'a}ndez-Cabrera, Juan F. and Hern{\'a}ndez-Su{\'a}rez, Elvio and Joven-{\'A}lvarez, Enrique and Otani, Chiko and Rubi{\~n}o-Mart{\'\i}n, J. Alberto",
    collaboration = "DALI",
    title = "{Discovery prospects with the Dark-photons {\&} Axion-like particles Interferometer}",
    eprint = "2303.03997",
    archivePrefix = "arXiv",
    primaryClass = "hep-ph",
    doi = "10.1103/PhysRevD.109.062002",
    journal = "Phys. Rev. D",
    volume = "109",
    number = "6",
    pages = "062002",
    year = "2024"
}

@article{Ringwald:2020ist,
    author = {Ringwald, Andreas and Sch{\"u}tte-Engel, Jan and Tamarit, Carlos},
    title = "{Gravitational Waves as a Big Bang Thermometer}",
    eprint = "2011.04731",
    archivePrefix = "arXiv",
    primaryClass = "hep-ph",
    reportNumber = "DESY 20-187, DESY-20-187, TUM-HEP-1293-20",
    doi = "10.1088/1475-7516/2021/03/054",
    journal = "JCAP",
    volume = "03",
    pages = "054",
    year = "2021"
}

@article{Domcke:2023bat,
    author = "Domcke, Valerie and Garcia-Cely, Camilo and Lee, Sung Mook and Rodd, Nicholas L.",
    title = "{Symmetries and selection rules: optimising axion haloscopes for Gravitational Wave searches}",
    eprint = "2306.03125",
    archivePrefix = "arXiv",
    primaryClass = "hep-ph",
    reportNumber = "CERN-TH-2023-093",
    doi = "10.1007/JHEP03(2024)128",
    journal = "JHEP",
    volume = "03",
    pages = "128",
    year = "2024"
}

@article{Graham:2025gtd,
    author = "Graham, Peter W. and Ramani, Harikrishnan and Simon, Olivier and Tanin, Erwin H.",
    title = "{Cosmological Limits on Strong Dark Forces}",
    eprint = "2511.09614",
    archivePrefix = "arXiv",
    primaryClass = "hep-ph",
    month = "11",
    year = "2025"
}

\end{document}